# Resurfaced CsPbBr$_3$ Nanocrystals Enable Free Radical Thermal Polymerization of Efficient Ultrafast Polyvinyl Styrene Nanocomposite Scintillators

Francesco Carulli[1*], Andrea Erroi[1], Francesco Bruni[1], Matteo L. Zaffalon[1], Mingming Liu[2], Roberta Pascazio[3,4], Abdessamad El Adel[5], Federico Catalano[6], Alessia Cemmi[7], Ilaria Di Sarcina[7], Francesca Rossi[8], Laura Lazzarini[8], Daniela E. Manno[9], Ivan Infante[5,10], Liang Li[11], and Sergio Brovelli[1*]

[1] *Department of Materials Science, University of Milano-Bicocca, via Roberto Cozzi 55, 20125 Milano, Italy*
[2] *School of Environmental Science and Engineering, Shanghai Jiao Tong University, Weiliu Rd., Minhang District, 200240, Shanghai P. R. China*
[3] *Materials Sciences Division, Lawrence Berkeley National Laboratory,1 Cyclotron Road, Berkeley, 94720, CA, USA*
[4] *Department of Materials Science and Engineering, University of California Berkeley, 210 Hearst Memorial Mining Building, 94720, CA, USA*
[5] *BCMaterials, Basque Center for Materials, Applications, and Nanostructures, UPV/EHU Science Park, Leioa 48940, Spain*
[6] *Istituto Italiano di Tecnologia, Via Morego 30, 16163, Genova, Italy*
[7] *ENEA Fusion and Technology for Nuclear Safety and Security Department, Via Anguillarese 301, Casaccia R.C. Roma, Italy*
[8] *IMEM-CNR Institute, Parco Area delle Scienze 37/A, 43124, Parma, Italy*
[9] *University of Salento, Dipartimento di Scienze e Tecnologie Biologiche ed Ambientali, Strada Provinciale 6, 73100, Lecce Italy*
[10] *Ikerbasque Basque Foundation for Science, Plaza Euskadi 5, Bilbao 48009, Spain*
[11] *Macao Institute of Materials Science and Engineering (MIMSE), Macau University of Science and Technology, Taipa, Macao, 999078 P. R. China.*
*francesco.carulli@unimib.it
*sergio.brovelli@unimib.it

**Abstract**

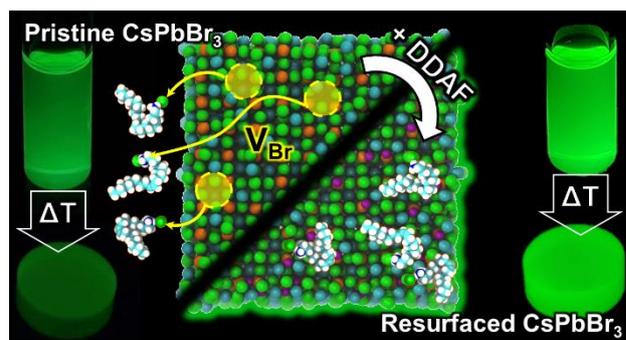

Lead halide perovskite nanocrystals (LHP-NCs) embedded in a plastic matrix are highly promising for a variety of photonic technologies and are quickly gaining attention as ultrafast, radiation-resistant nanoscintillators for radiation detection. However, advancements in LHP-NC-based photonics are hindered by their well-known thermal instability, which makes them unsuitable for industrial thermally activated mass polymerization processes - crucial for creating polystyrene-based scintillating nanocomposites. In this study, we address this challenge by presenting the first thermal nanocomposite scintillators made from CsPbBr$_3$ NCs passivated with fluorinated ligands that remain attached to the particles surfaces even at high temperatures, enabling their integration into mass-cured polyvinyl toluene without compromising optical properties. Consequently, these nanocomposites demonstrate scintillation light yields reaching 10,400 photons/MeV, sub-nanosecond scintillation kinetics, and remarkable radiation resilience, able to withstand gamma radiation doses of up to 1 MGy. This performance not only meets but also exceeds the scintillation of plastic scintillators, despite the radiation-induced damage to the host matrix.

Radiation detectors are at the core of several strategic technologies with profound scientific, societal and industrial impact[1], such as nuclear monitoring in factories[2], power plants[3,4] and for borders patrol [5,6], in industry, medical therapy[7,8], diagnostics[9-11], environmental control[12], and in frontier experiments in high energy and particle physics[1,13,14]. All of these technologies require a high probability of interaction with ionizing radiation[15], prompted by the use of high atomic number (Z) compounds, high scintillation efficiency (Light Yield, LY) representing the number of photons radiated per unity of deposited energy), and fast emission kinetics. The latter is crucial when operating in the so-called time-of-flight (ToF) mode[16], adding the time coordinate to the detection capability and underpinning applications such as ToF positron emission tomography (ToF-PET) for high precision diagnostics[17-19] and suppressed pile up calorimeters for high-luminosity hadron colliders [20, 21].

To overcome the cost and scalability limitations of conventional scintillator crystals and the low stopping power and radiation stability (i.e., low radiation hardness[22,23]) of plastic scintillators, the use of high-Z colloidal inorganic nanocrystals (NCs) as nanoscintillators embedded in polymers is receiving increasing attention[24-29]. Indeed, the integration of scintillating NCs in plastic waveguides can offer the best of both approaches, allowing scalable and low-cost fabrication of highly emissive materials with flexible shape and size[25, 30] and potentially largely increased radiation hardness[31] and scintillation timing capability thanks to the activation of quantum-size effects, such as the recently demonstrated ultrafast scintillation of stable multiexcitons created following interaction with ionizing radiation[27, 32, 33].

This technological potential has led to the exploration of various combinations of polymers and NCs (including chalcogenides[34, 35], oxides[36, 37] and halides of heavy metals[38]), almost importantly all-inorganic lead halide perovskite (LHP) NCs, such as $CsPbBr_3$[24, 29, 39-41]. These systems are prized for their unique ability to be synthesized in large batches at room temperature[42-44], high exciton and biexciton binding energy[45, 46] resulting in a strong multi-excitonic decay at room temperature[32, 33, 47] high tolerance to structural defects[48, 49] and exceptional radiation hardness resulting in stable and highly efficient scintillation[31, 50, 51].

Despite these notable advantages, a major challenge arises from the inherent susceptibility of LHP NCs to thermal degradation[52-54], which limits their compatibility with the most established polymerization protocols for scintillating polymers such as polystyrene (PS) or polyvinyl toluene (PVT), which constitute the body of high-performance plastic scintillators and actively contribute to their scintillation emission both by direct emission and by sensitizing the luminescence of embedded dyes[25, 55, 56]. Unlike polyacrylates, such as polymethylmethacrylate (PMMA), which can be photopolymerized at room temperature[57-60], but lack scintillation properties and may damage the incorporated NCs due to partial polarity[61, 62], PS derivatives preserve the optical properties of the NCs. However, they require a thermal budget above 60°C to overcome the radical stabilization caused by the aromatic rings, which hinders the breaking of the vinyl double bond and the initiation of radical polymerization[63, 64]. This quite harsh condition results in the degradation of the particle surfaces and in phase transition to non-emissive allotropes[48, 65], with a consequent drop in the NC emission efficiency.

Although there have been many recent advances aimed at maintaining the optical properties of NCs at elevated temperatures[66-69], the encapsulation of NCs in PS has so far been addressed by proof-of-concept studies based on NCs in PS matrices obtained by evaporation of a common solvent[47, 70, 71], which are not suitable for technological applications, or on photopolymerized composites of polyacrylates derivatives such as poly lauryl-methacrylate which have poorer optical properties and lack scintillation capability[32, 33, 58]. As a result, the interplay between NCs and a scintillating host is still largely unexplored and unexploited. We emphasize that such a processing limitation of NCs via temperature-driven processes is not only critical for radiation detection but is a general issue for optical/photonic technologies that require highly emissive nanocomposites, such as displays or artificial lighting sources. Therefore, devising strategies for compatibilizing LHP NCs with the thermal polymerization could offer a general solution for many photon-management applications using NCs emitters.

Here, we aim to contribute to this endeavor by demonstrating the first example of mass polymerized PVT nanocomposites embedding $CsPbBr_3$ NCs that retain their original optical properties. This substantial advance is made possible by exploiting the surface passivation of $CsPbBr_3$ NCs with didodecyl-dimethylammonium fluorinated ligands, which have recently been shown to enable anti-thermal quenching of NCs photoluminescence (PL). Our experiments, corroborated by classical molecular dynamics (MD) calculation, revealed that the replacement of native oleylammonium/oleate ligands with DDAF moieties protects the NCs from thermal annealing at 80°C, which largely damages untreated control particles due to massive ligand detachment, and allowed us to fabricate a set of highly emissive nanocomposites with NCs mass concentration up to [NC]=10 wt.%, retaining over 90% of the initial luminescence yield. This enabled us to interrogate the photophysical interaction between the scintillating polymer matrix and the embedded NCs, as well as the effects on the scintillation timing and the radiation hardness of the nanocomposite material. We observed that the introduction of increasing amounts of $CsPbBr_3$ NCs progressively turns the UV scintillation of the PVT host into the green NCs luminescence through a combination of radiative and nonradiative energy transfer processes. For [NC]>0.5 wt.%, the direct interaction of NCs with ionizing radiation becomes dominant and enhances the light output by over an order of magnitude, reaching a light yield of LY=10400 ph/MeV and adding an ultrafast (<120 ps) multiexcitonic contribution to the scintillation dynamics. This substantially reduces the average scintillation lifetime to 1.1 ns, which is very promising for fast timing applications. Moreover, systematic monitoring of nanocomposite exposed to cumulative $^{60}$Co gamma doses up to 1 MGy shows that for [NC]>2 wt.% the NCs scintillation is completely unaffected by the radiation damage suffered by the polymer matrix. The fully preserved pre-irradiation performance makes them promising candidates for operations in harsh radioactive environments. These results address the limitations of using LHP NCs in plastic matrices cured by industrial techniques, paving the way for high-performance nanocomposite scintillators through the combination of scintillating polymer matrices and perovskite NCs.

***Synthesis and resurfacing of CsPbBr$_3$ NCs.*** Oleate/oleylammonium-capped CsPbBr$_3$ NCs (lateral size $d$=7.0±0.6 nm) were synthesized via a high-temperature hot-injection method and then subjected to a ligand exchange procedure with DDAF precursor at room temperature[31,72] (described in **Methods**).

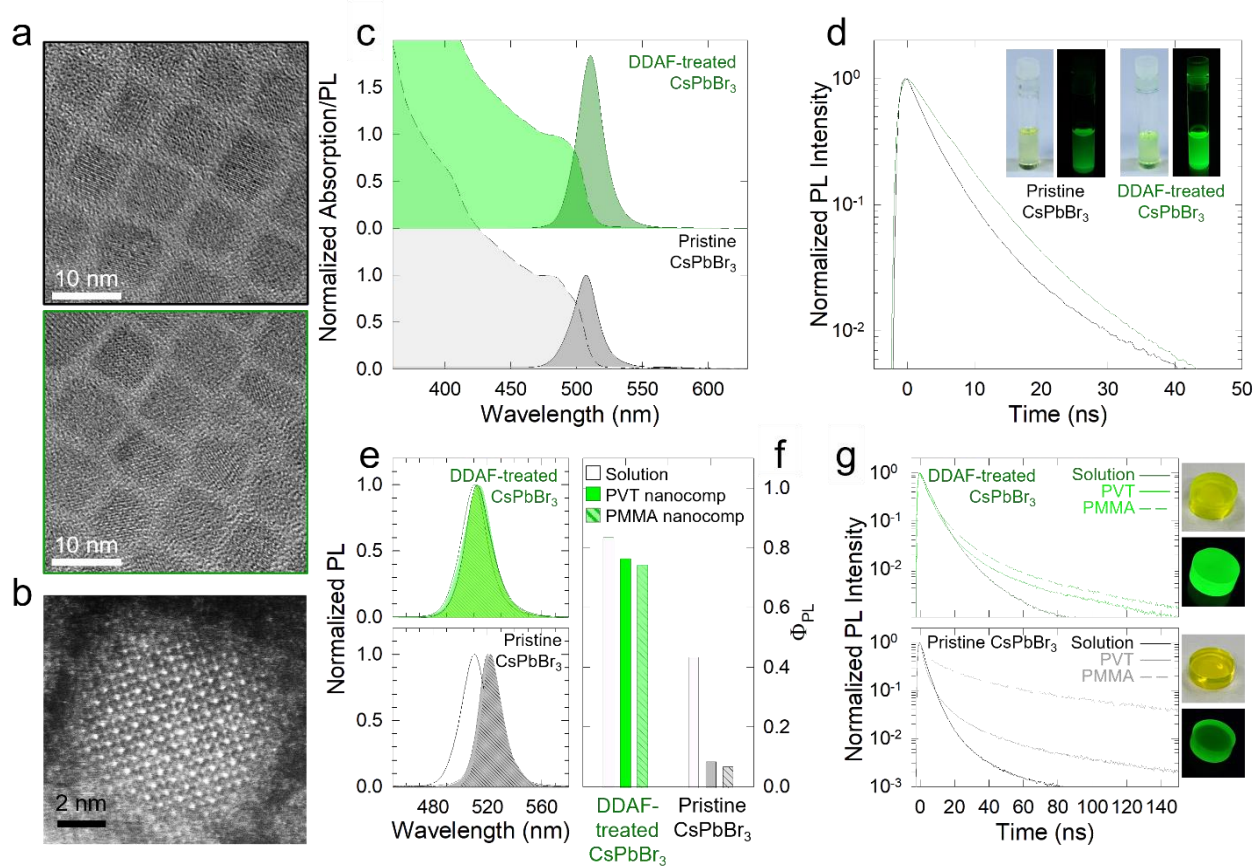

**Figure 1**: a) TEM images of pristine (top) and DDAF-treated CsPbBr$_3$ (bottom) NCs. b) Enlargement of a single DDAF-treated CsPbBr$_3$ NC obtained via aberration-corrected STEM-HAADF. c) Absorption (dashed lines) and PL (continuous

lines) spectra of pristine and DDAF-treated CsPbBr$_3$ NCs. d) PL decay curves of the same samples excited at 405 nm. Photographs of the colloidal solutions under room light and 365 nm illumination are shown in the inset. e) Comparison between PL spectra of DDAF-treated and pristine NCs in solution (lines), in PVT (shaded areas) and in PMMA (patterned areas) nanocomposites and respective f) $\Phi_{PL}$-values and g) normalized PL decay curves. All measurements were excited at 405 nm. The same color code applies throughout the whole figure.

The resurfacing treatment did not cause any morphological or structural changes in the NCs as demonstrated by the TEM images of CsPbBr$_3$ NCs before (hereafter referred to as pristine CsPbBr$_3$) and after DDAF treatment reported in **Figure 1a**. The comparison between the diffraction patterns before and after the treatment and the NCs size distribution are reported in **Supporting Figure S1**. Energy-dispersive X-ray spectroscopy maps (EDX) of the DDAF-treated samples confirm that fluorine is present and homogeneously distributed throughout the NCs (**Supporting Figure S2**). The lattice spacings measurements were performed on tens of individual crystals, of which a typical STEM-HAADF image is reported in **Figure 1b**. Within the resolution limit of the technique, no significant differences were detected among the different parts of the DDAF-treated CsPbBr$_3$ NCs (shown in **Supporting Figure S3**), which indicates that fluorine is located only on the surfaces of the NCs. Consistently, the analysis of the optical properties of pristine and DDAF-treated NCs in toluene (**Figure 1c**) indicates that the treatment does not alter their electronic structure, as highlighted by the comparison of the PL and absorption spectra, which show no changes in spectral shape and position (both centered at 510 nm with full width half maximum, FWHM=22.5 nm). Crucially, the DDAF treatment boosts the PL quantum yield from $\Phi_{PL}$=45±3% to $\Phi_{PL}$=83±5% (full spectra in **Supporting Figure S4**) and leads to slower initial PL dynamics compared to the pristine NCs (**Figure 1d**), confirming the suppression of non-radiative traps associated with surface (halide) vacancies induced by the displacement of weakly binding oleate/oleylammonium ligand pairs[31, 73-75].

***Thermally polymerized CsPbBr$_3$-based nanocomposites.*** Based on these promising results, we proceeded to incorporating the CsPbBr$_3$ NCs into polymer nanocomposites and to studying their optical properties. To generalize the approach, we chose PVT and PMMA as scintillating and non-scintillating matrixes. Dried NCs were dissolved in a monomer mixture (4-methylstyrene:divinylbenzene 95:5% V/V for PVT or methyl methacrylate for PMMA) with 1000 ppm of dilauroyl peroxide as thermal initiator via mild sonication, subsequently poured in cylindrical molds sealed in an inert atmosphere and finally thermally polymerized in a thermal bath at 65°C for 24 h. Due to the increased resistance to high temperatures induced by DDAF treatment, both PVT and PMMA nanocomposites based on these NCs show negligible changes in their optical properties, as confirmed by the preservation of the PL spectrum (**Figure 1e**) and $\Phi_{PL}$=76% and 73%, corresponding to over 90% of the initial value. Consistently, the PL decay trace is largely unchanged except for the appearance of a weak slow tail (<10% relative weight) ascribed to trapping in shallow surface defects[31, 61, 76] possibly resulting from the minor detachment of surface ligands due to the high radical reactivity at these temperatures[77]. Crucially, nanocomposites prepared by the same protocol using pristine NCs suffer from major deterioration of their optical properties, with a drop in $\Phi_{PL}$ of over 80%-90%, accompanied by an acceleration of the initial PL dynamics due to increased trapping losses, a marked spectral redshift and a more pronounced slow decay tail ascribed to the recombination of shallowly trapped excitons[31, 61, 76] that is largely dominant in PMMA which causes stronger PL drop likely due to its polar nature that further deteriorates the NCs (**Figure 1e-g**). To better understand the origin of the higher thermal stability of DDAF-treated NCs, we conducted a classical molecular dynamics (MD) study on the DDAF-capped NCs and on the pristine NCs with oleate and oleylammonium as native ligands[72] (see **Supporting Information** for details). **Figure 2a** reports the statistical analysis of the distances between the average position of the anchoring group of the ligands and the center-of-mass of the NC during the MD simulations at room temperature and at 65°C. The analysis was carried out by monitoring an equilibrated 50 ns long trajectory.

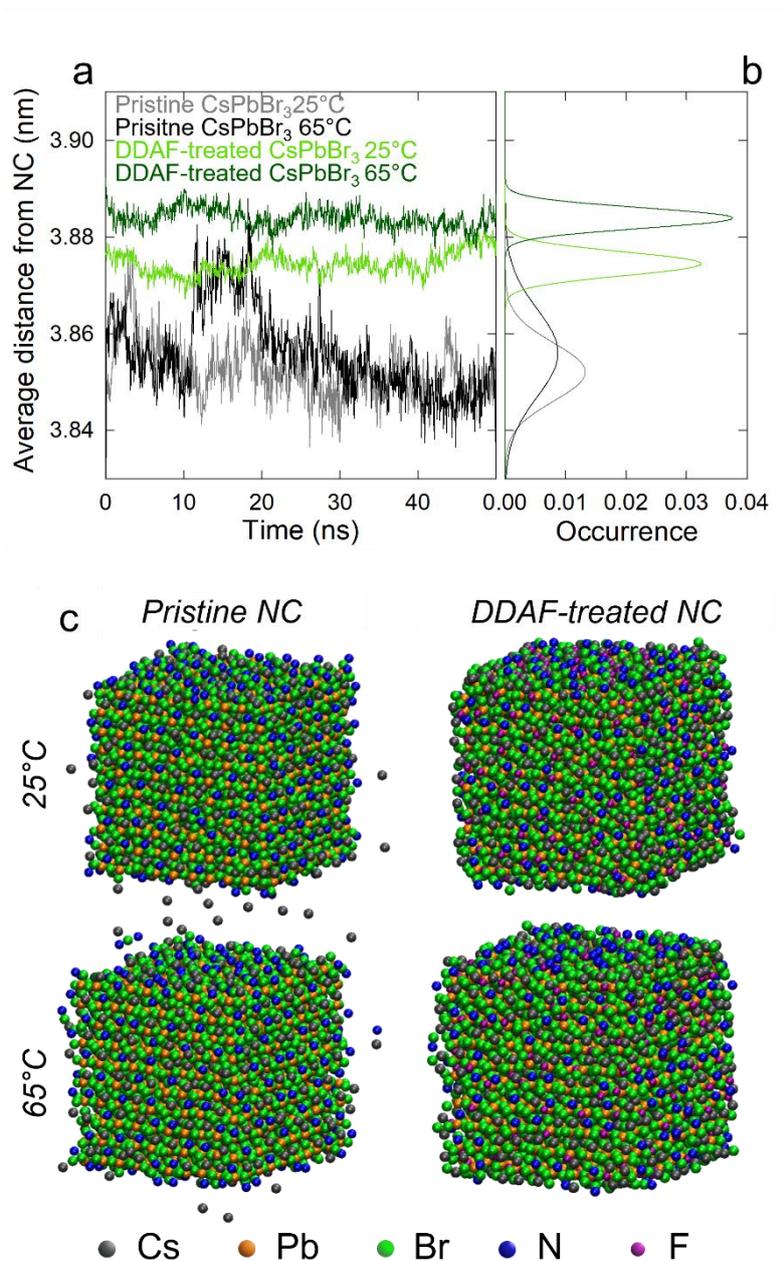

**Figure 2**: a) Time evolution of the average distances between the anchoring group of the ligands and the center of mass of the NC of pristine and DDAF-treated CsPbBr$_3$ NCs at 25°C (grey and light green curves) and 65°C (black and dark green curves, details of the simulation are reported in **Supporting Information**). The average distance was calculated over 220 ligands. b) Time-mediated statistic of the data shown in 'a'. c) Frames extracted from a molecular dynamics simulation of pristine (left) and DDAF-treated (right) CsPbBr$_3$ NC at 25°C (top) and 65°C (bottom). The frames show only the nitrogen and fluorine atoms of the ligand anchoring group and NCs lead and cesium atoms previously belonging to the NC structure torn from the surface as a result of the ligand detachment.

As highlighted by the time-mediated statistics shown in **Figure 2b**, DDAF ligands show minimal broadening around their initial position, which is consistent with stable capping independent of temperature. In contrast, the distance distribution of the oleylamine ligands in the pristine NCs show a significantly larger standard deviation, which suggests a lower stability potentially leading to partial detachment of the ligands from the NC surface. This is more pronounced for oleylammonium than oleate, as the latter is bound directly to underneath Pb atoms (**Supplementary Figure S5**). These findings are in accordance with the lower $\Phi_{PL}$ of pristine NCs compared to DDAF-treated particles and with the previously reported relation between the presence of surface halide undercoordination and PL decay dynamics in analogue NCs[31, 61, 72]. More importantly, when evaluating this mechanism at 65°C (the temperature of the polymerization of PVT-based nanocomposites) the behavior of the surface ligands is more divergent, with the DDAF showing stability similar to that observed at room temperature and pristine ligands suffering from an even more pronounced detachment, once again in accordance with the degradation of the optical properties following the nanocomposite fabrication. These results are corroborated also by the root mean square deviation (RMSD) analysis, which evaluates the ligands detachment by comparing their positional deviation over time with respect to a reference position chosen as time zero of the MD simulation (**Supporting Figure S6**). Also in this case, the DDAF ligands remain stable on the NC surface, while the oleylamine was found to detach as an oleylammonium-bromide complex, removing bromide ions from the surface of the NC, in accordance with the previously described mechanism[78-80]. The oleylammonium-bromide complex could also undergo a neutralization process, not captured in the MD simulation, that would favor the formation of HBr and neutral oleylamine, which would render the detachment process essentially irreversible.

***Scintillation properties of thermally polymerized nanocomposites.*** Based on the remarkable resistance of DDAF-treated $CsPbBr_3$ NCs (henceforth referred to simply as NCs) to the thermal polymerization processes, we proceeded with the fabrication of a set of nanocomposites with identical size and geometry containing progressively higher [NC], ranging from 0.01 wt.% to 10 wt.%. Having established that the optical properties of the NCs embedded in the matrix are independent of their concentration (as reported by the PLQY vs. [NC] analysis shown in **Supporting Fig. S7**), we proceed to investigate the scintillation properties of the nanocomposites. The sample size was chosen to completely attenuate the incident ionizing radiation in all cases (see **Supporting Information** for the detailed calculation), so that the evolution of the scintillation properties with [NC] is due to the photophysics of the composites and not to the variation of the total stopping power of impinging X-rays. The RL spectra of the nanocomposites measured in the same collection geometry under identical continuous X-ray excitation together with a bare PVT control sample are shown in **Figure 3a**.

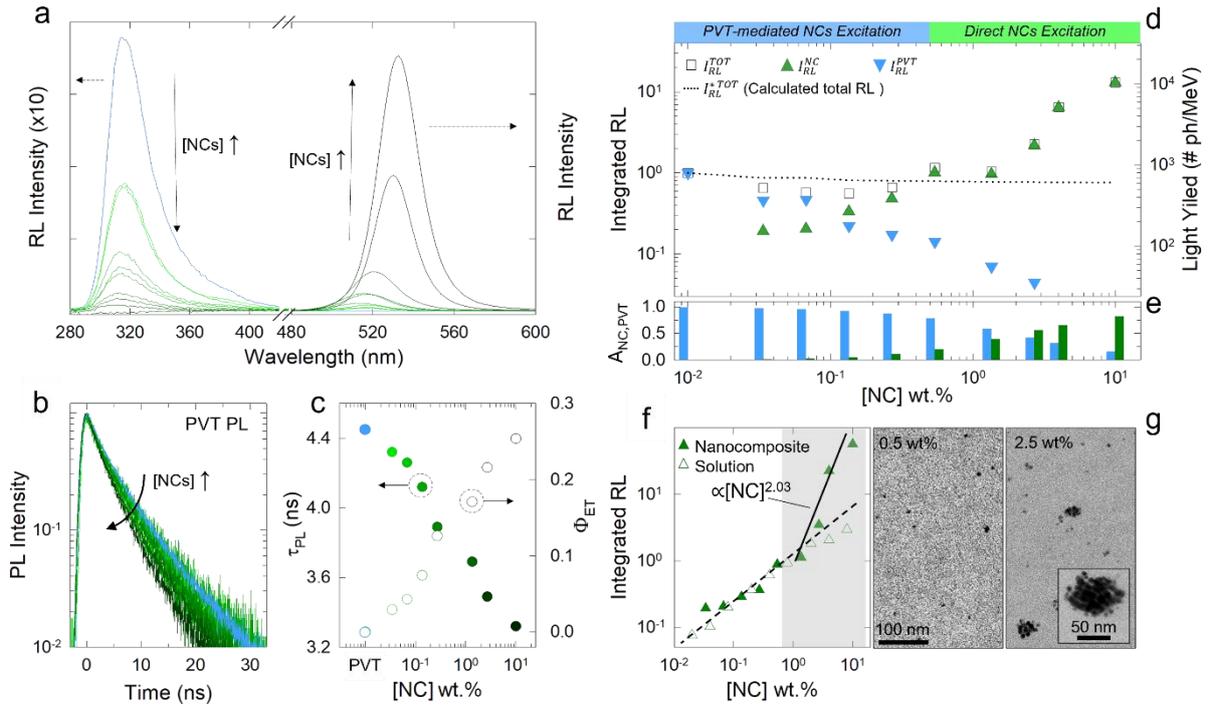

**Figure 3:** a) RL spectra of nanocomposites with increasing [NC] from 0.01 to 10 wt.% collected under X-ray continuous irradiation (20 kV, 20 mA). The blue line is the RL of a bare PVT sample. Notice that the intensity of the PVT RL from all samples has been scaled by a factor 10. b) Decay curves of the PVT emission at 310 nm for the same samples as in 'a' (increasing [NC] from light to dark green) under 250 nm excitation. c) Effective PL lifetime of the PVT emission (filled circles) extracted from the dynamics in 'b' (as the time after which the intensity has dropped by a factor $e$) and corresponding energy transfer efficiency (hollow circles). d) Spectrally integrated total RL intensity ($I_{RL}^{TOT}$ white squares) together with the relative integrated contributions of PVT ($I_{RL}^{PVT}$, blue triangle) and NCs ($I_{RL}^{NC}$, green triangle) extracted from the spectra in 'a'. e) Relative contribution by PVT matrix (blue bars) and NCs (green bars) to the total mass attenuation coefficient as a function of [NC]. f) Reabsorption-corrected integrated RL intensity of CsPbBr$_3$ NCs in nanocomposites (filled triangles) and colloidal suspensions in octane (hollow triangles) as a function of [NC]. g) TEM images of 70 nm thick sheet of nanocomposites with [NC]=0.5 and 2.5 wt.%. In the inset the magnification of a NCs clusters is reported.

The same experimental conditions were also used to estimate the LY by side-by-side comparison with a commercial plastic scintillator (EJ276, LY=8600 ph/MeV) with same dimensions as the nanocomposites. The pure PVT sample shows its characteristic RL spectrum at 310 nm. The same UV emission is also present in the nanocomposites showing a gradual decrease with increasing [NC] due to the combined effect of reabsorption and energy transfer to the NCs (vide infra), accompanied by the progressive intensification of the RL of the CsPbBr$_3$ NCs. To clarify the sensibilization mechanism we measured the time dynamics of the PVT-PL as a function of [NC]. As shown in **Figure 3b**, the time dynamics is essentially single exponential in all samples and undergoes mild acceleration from $\tau_{EFF}$=4.4 ns (calculated as the time after which the intensity has decreased by a factor $e$) to $\tau_{EFF}$=3.3 ns at the highest [NC] value, corresponding to a maximum energy transfer efficiency, $\Phi_{ET} = 1 - \tau_{PVT}([NC])/\tau_{PVT}$ =25% (**Figure 3c**), suggesting the occurrence of nonradiative energy transfer (Forster or Dexter alike), yet insufficient to justify the complete disappearance of the PVT emission.

This points to a major role of radiative NC excitation by absorption and re-emission of the PVT scintillation as further confirmed by the radiation hardness measurements reported later in this work. We also notice the gradual redshift of the NCs RL peak with increasing [NC] due to increasing reabsorption of the NC scintillation by the NCs themselves, in accordance with the partial spectral overlap between their absorption onset and emission peak. The trends of the relative contribution of the PVT and the NCs emission to the integrated RL intensity (namely $I_{RL}^{PVT}$, $I_{RL}^{NC}$ and $I_{RL}^{TOT}$) are reported in **Figure 3d** together with the corresponding total LY values. **Figure 3e** shows the concomitant evolution of the relative contributions ($A_{PVT}$ and $A_{NC}$) of the PVT matrix (with attenuation coefficient $(\mu/\rho)_{PVT}$) and the NCs (with attenuation coefficient $(\mu/\rho)_{NCs}$)) to the total mass attenuation coefficient of each nanocomposite expressed as $A_{PVT,NC} = \frac{(\mu/\rho)_{PVT,NCs}}{(\mu/\rho)_{PVT}+(\mu/\rho)_{NCs}}$, which provides information about the branching ratio of energy deposited in the PVT and in the NCs. For [NC]<0.5 wt.%, where the attenuation is dominated by PVT ($A_{PVT}$>$A_{NC}$), and output photons are mostly generated by PVT scintillation and its down-conversion into the NCs emission, the total integrated RL intensity is nearly constant except for a slight reduction compared to bare PVT (with intensity $I_{RL}^{PVT}(0)$) due to the non-unity $\Phi_{PL}$ of the NCs (76%). Consistently the RL intensity in this regime can be approximated by the expression $I_{RL}^{*TOT}([NC]) = I_{RL}^{PVT}([NC]) + \left(1 - \frac{I_{RL}^{PVT}([NC])}{I_{RL}^{PVT}(0)}\right)\Phi_{PL}$ which neglects non radiative energy transfer for simplicity. Notably, increasing [NC]>0.5 wt.% results in a steep increase of $I_{RL}^{TOT}$, reaching a corresponding LY=10400 ph/MeV at 10 wt.%, which is higher than commercial plastic scintillators such as EJ276 (LY=8600 ph/MeV) or EJ208 (9200 ph/MeV). This steep growth in the total RL intensity is dominated by $I_{RL}^{NC}$ and accompanied by an increase of $A_{NCs}$, which makes direct interaction with ionizing radiation the main NC excitation source ($A_{NC}$>25% for [NC]>0.5 wt.%). Interestingly, the $I_{RL}^{NC}$ vs. [NC] trend becomes quadratic after correcting for reabsorption losses (**Figure 3f** and **Supporting Figure S8**). In order to better understand this trend, we performed a comparative RL study of nanocomposites and colloidal NCs solutions in octane (a non-aromatic, non-scintillating solvent with X-ray attenuation coefficient similar to PVT) as a function of [NC]. **Figure 3f** reports the integrated RL intensity corrected for the reabsorption losses of the two sets: the colloidal solution exhibits the same linear dependence on [NC] throughout the whole investigated range of concentration, indicating that in a homogeneous dispersion and in the absence of a scintillating host, the NCs are subject to only one excitation process and that the capability of interact with ionizing radiation scales with the NC amount. Conversely, the nanocomposites show two different dependences of $I_{RL}^{NC}$ with [NC]: at low loadings, the trend is nearly linear as for the solution, whereas $I_{RL}^{NC} \propto [NC]^{2.03}$ for [NC]>0.5 wt.%, which indicates that two distinct excitation processes concurring in generating the NC's scintillation are active only in the nanocomposites. One possible explanation for this is that in the high loading regime, the NCs not only emit light by direct interaction with X-ray excitation (with a minor aid by down-converting by downshifting the PVT emission) but also might act as secondary excitation sources for other NCs through the release of photoelectrons inside the plastic matrix. Recent studies have indeed shown that NCs retain a very minor part (less than 1%) of the incident energy under soft X-ray excitation, whereas most photoelectrons escape the NCs[81,82]. The different behavior between the colloidal systems and the nanocomposites could thus be due to different local arrangement of the NCs in the latter that promotes interparticle secondary excitation. This scenario is consistent with the TEM images of 70 nm thin slices of PVT composites containing [NC]=0.5 and 2.5 wt.% reported in **Figure 3g** (TEM of nanocomposite with [NC]=10 wt. % is reported in **Supplementary Figure S9**) showing that in low-[NC] nanocomposites, NCs distribute evenly inside the matrix while for high particle contents they tend to accumulate into NCs-rich domains. In depth study of interparticle interactions are however beyond the scope of this work and will be treated with highly controlled systems in a separate study. Crucially the transition from the "PVT-mediated" to the "direct" scintillation regime not only enhances the light output but substantially accelerates the scintillation kinetics by prompting ultrafast multi-excitonic emission of the NCs.

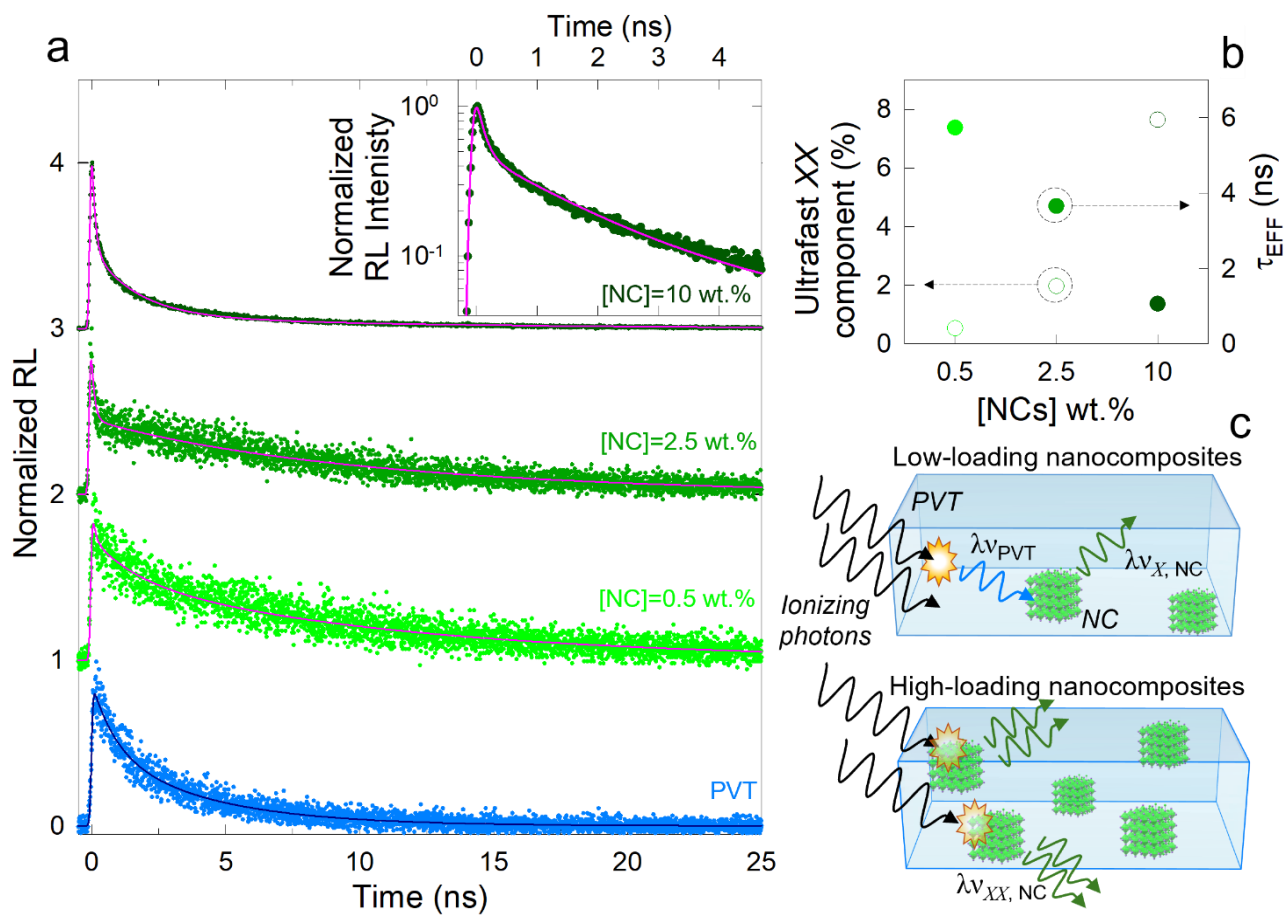

**Figure 4**: a) Scintillation decay of bare PVT and nanocomposites with [NC]= 0.5, 2.5 and 10 wt.%. The continuous dark blue and pink curves are the fit functions. The scintillation decay of the 10 wt.% sample in logarithmic scale is shown the inset. b) Weight of the prompt component (hollow circles) and effective recombination times $\tau_{EFF}$ (filled circles) extracted from the fitting of the decay traces shown in plot a. c) Sketch of the scintillating mechanism involved in the scintillation process in nanocomposite with low (top scheme) and high (bottom scheme) [NC]. $\lambda\nu_{PVT}$, $\lambda\nu_{X, NC}$ and $\lambda\nu_{XX, NC}$ refer to PVT scintillation, NC emission via single exciton recombination and NC emission via multiexciton recombination, respectively.

**Figure 4a** shows the RL kinetics of a pure PVT sample and three nanocomposites with [NC]=0.5, 2.5 and 10 wt.%. The decay traces were fitted with exponential functions convoluted with the IRF of our detection chain. The PVT sample shows an essentially single exponential decay with 4.0 ns lifetime matching the corresponding PL (see **Figure 3b** and **Supporting Figure S10**). The nanocomposites, on the other hand, feature three decay components with relative weights that depend strongly on [NC]: $\tau_X$=10 ns matching to the PL lifetime of the NCs and ascribed to single exciton ($X$) decay, a faster $\tau_{X*}$=1.6 ns, contribution ascribed to charged NC excitons ($X^*$) and an ultrafast component $\tau_{XX}$=120 ps lifetime due to the recombination of multi-excitons generated by X-ray excitation, in agreement with previous reports[32, 33, 47, 61]. Also consistent with the cw-RL in **Figure 3d**, the main kinetic component of the composite with [NC]=0.5 wt.% is the slow emission, confirming that dominant process in the low concentration regime is the downshifting of the PVT scintillation, as schematized in the **Figure 4c**. As [NC] increases, the weight of the ultrafast multi-excitonic component grows steeply, as reported in **Figure 4b**, reaching a relative weight as high as 8% for the 10 wt.% nanocomposite. This results in a substantial acceleration of the effective decay time $\tau_{EFF}$ - calculated as the harmonic average of the decay components weighted for their respective integrated intensity (see Supporting Information for details) – from 6 ns to 1.1 ns in the most concentrated case (**Figure 4b**) which is substantially faster than PVT. From these parameters it is interesting to estimate the potential coincidence time resolution (*CTR*) obtainable with these nanocomposites for ToF application using the formula: $CTR = 3.33\sqrt{\frac{\tau_{RISE} \cdot \tau_{EFF}}{N}}$, where $N$ is the estimated number of emitted photons for 511 keV excitation and $\tau_{RISE}$ is the 10-90% risetime, set here to 90 ps. For the highest concentrated nanocomposite, we obtained an estimated CTR as fast as 14 ps, which is promising for fast timing applications.

***Radiation stability of CsPbBr$_3$-PVT nanocomposites.*** Alongside the remarkable scintillation and timing performance, the use of nanocomposites based on NCs also has considerable advantages in terms of resistance to prolonged exposure to ionizing radiation that is particularly important for use in harsh radioactive environments as well as in high luminosity colliders. The radiation hardness of these nanocomposites was studied by monitoring the RL before and after irradiation with $^{60}$Co gamma rays (1 MGy of total irradiated dose), which was carried out at the calibrated irradiation facility Calliope[83] of the ENEA Fusion and Technology for Nuclear Safety and Security Department. As shown by the photograph in the inset of **Figure 5a**, the PVT control samples was highly damaged by irradiation leading to macroscopic color change from transparent to yellow and to the rise of a strong absorption shoulder below 450 nm (shaded area in **Figure 5a**) due to the breakdown of the polymer chains into UV absorbing oligomeric/molecular subunits[22, 23, 84].

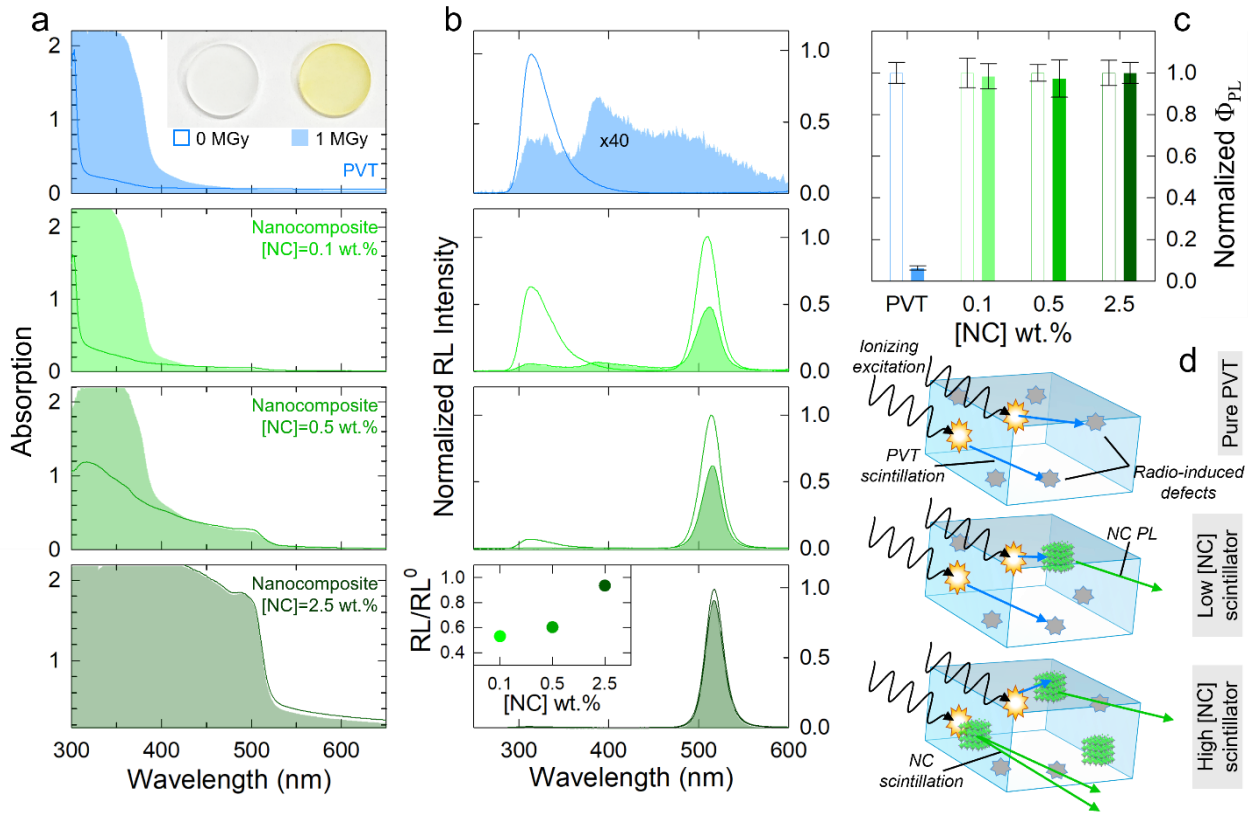

**Figure 5:** a) Absorption spectra of bare PVT and nanocomposite with different [NC] (0.1, 0.5 and 2.5 wt.%) before (hollow spectra) and after (shaded spectra) exposition to 1 MGy $^{60}$Co gamma ray radiation. The photos of a bare PVT scintillator before and after exposition to the same dose are shown in the inset. b) RL spectra of the same set of samples reported in panel a, before (hollow spectra) and after (shaded spectra) exposition to 1 MGy $^{60}$Co gamma ray radiation. The RL spectrum of bare PVT was enlarged by a factor 40 for clarity. In the inset we reported the ratio between RL measured before and after exposition to 1 MGy $^{60}$Co gamma ray radiation for the same nanocomposites sample set. c) Normalized $\Phi_{PL}$ intensity of the same set of samples shown in panel a, before (hollow bars) and after (filled bars) exposition to 1 MGy gamma ray radiation. d) Sketch of the competitive absorption phenomenon due to radio-induced defects of PVT in a pure PVT scintillator and in a highly concentrated scintillators based on CsPbBr$_3$ NCs.

Such radio-induced defects result in nearly complete reabsorption of the PVT scintillation (**Figure 5b**). Similar modification of the absorption spectrum is experienced by the nanocomposites, leading to analogous suppression of $I_{RL}^{PVT}$ and partial dimming of $I_{RL}^{NC}$ of the low-[NC] samples. Crucially, as shown in **Figure 5c**, $\Phi_{PL}$ of the nanocomposites under direct excitation of the NCs at 470 nm is essentially unchanged after irradiation, confirming high radiation hardness of CsPbBr$_3$ NCs[31] also in PVT matrices (full spectra are reported in **Supporting Figure S11**). This further indicates that the reduction of $I_{RL}^{NC}$ in low-[NC] samples is due to competitive absorption by radio-induced defects of the PVT scintillation, which reduces the PVT-mediated excitation rate of the NCs, and not to radio-induced damaging of the NCs. Looking at the [NC]=0.1 wt.% case in **Figure 5b**, we further notice that the quenching of the NC emission is weaker than the corresponding PVT contribution (90% vs 50% reduction) which is likely a combined effect of direct NC excitation and the fact that in the nanocomposite some of the PVT-emitted photons are downshifted to 520 nm by the NCs before being intercepted and reabsorbed by the radio-induced defects of PVT. More importantly, as [NC] reaches the onset of the direct scintillation regime, radio-induced damage becomes irrelevant because the NC emitted photons at 520 nm are completely out of resonance with the radio-induced UV absorption band (as shown in **Figure 5d**). As a result, high-loading composites preserve over 95% of their initial RL intensity even after the extreme irradiated dose.

In conclusion, we have realized for the first time a polymeric nanocomposite containing $CsPbBr_3$ perovskite NCs through a thermal mass polymerization approach without incurring in the ubiquitous quenching of the NCs luminescence. This is ascribed to the high resistance of DDAF ligands to thermal detachment, thus preserving the original surface passivation. The study of the scintillation performance of the nanocomposites with increasing particles loading shows that the NC embedded in the scintillating matrix turn from wavelength shifters of the PVT scintillation to active nanoscintillators following direct interaction with ionizing radiation, resulting in ultrafast and efficient scintillators. The extreme radiation hardness of these NCs ensures that the scintillators maintain their performance even after prolonged exposure to radiation despite exhibiting damaging of the host matrix. This durability, combined with the straightforward preparation strategy and superior performance, positions these nanocomposite scintillators as promising candidates for advanced applications in radiation detection and fast timing technologies.

**Associated Content**

**Supporting Information**

Details of the experimental and computational methods. Calculated force field parameters, TEM images and size analysis of pristine and treated NCs, EDX maps of a DDAF-treated NCs samples, STEM-HAADF images and line profile intensity of DDAF-treated NCs. Comparative PL and RL spectra, PL quantum yield as a function of the NC content. TEM images of PVT composites containing [NC]=10 wt.% and RL spectra before and after gamma irradiation tests.

**Acknowledgments**

This work was funded by Horizon Europe EIC Pathfinder program through project 101098649 – UNICORN, by the PRIN program of the Italian Ministry of University and Research (IRONSIDE project), by the European Union—NextGenerationEU through the Italian Ministry of University and Research under PNRR—M4C2-I1.3 Project PE_00000019 "HEAL ITALIA", by European Union's Horizon 2020 Research and Innovation programme under Grant Agreement No 101004761 (AIDAINNOVA). This research is funded and

# References

1. Dujardin, C.; Auffray, E.; Bourret-Courchesne, E.; Dorenbos, P.; Lecoq, P.; Nikl, M.; Vasil'ev, A. N.; Yoshikawa, A.; Zhu, R. Y., Needs, Trends, and Advances in Inorganic Scintillators. *IEEE Trans. Nucl. Sci.* **2018,** *65* (8), 1977-1997.
2. Jenkins, D., Radiation Detection for Nuclear Physics. In *Methods and industrial applications* [Online] IOP Publishing: 2020. https://dx.doi.org/10.1088/978-0-7503-1428-2.
3. Belousov, M. P.; Gromyko, M. V.; Ignatyev, O. V., Scintillation γ spectrometers for use at nuclear power plants (review). *Instrum. Exp. Tech.* **2017,** *60* (1), 1-19.
4. Kim, C.; Lee, W.; Melis, A.; Elmughrabi, A.; Lee, K.; Park, C.; Yeom, J.-Y., A Review of Inorganic Scintillation Crystals for Extreme Environments. **2021,** *11* (6), 669.
5. Cutmore, N. G.; Liu, Y.; Tickner, J. R. In *Development and commercialization of a fast-neutron/x-ray Cargo Scanner*, 2010 IEEE International Conference on Technologies for Homeland Security (HST), 8-10 Nov. 2010; 2010; pp 330-336.
6. Glodo, J.; Wang, Y.; Shawgo, R.; Brecher, C.; Hawrami, R. H.; Tower, J.; Shah, K. S., New Developments in Scintillators for Security Applications. *Phys. Procedia* **2017,** *90*, 285-290.
7. Lecoq, P., Development of new scintillators for medical applications. *Nucl. Instrum. Methods Phys. Res. A.* **2016,** *809*, 130-139.
8. Lu, L.; Sun, M.; Lu, Q.; Wu, T.; Huang, B., High energy X-ray radiation sensitive scintillating materials for medical imaging, cancer diagnosis and therapy. *Nano Energy* **2021,** *79*, 105437.
9. Fredenberg, E., Spectral and dual-energy X-ray imaging for medical applications. *Nucl. Instrum. Methods Phys. Res. A.* **2018,** *878*, 74-87.
10. Ito, M.; Hong, S. J.; Lee, J. S., Positron emission tomography (PET) detectors with depth-of-interaction (DOI) capability. *Biomed. Eng. Lett.* **2011,** *1* (2), 70-81.
11. Hosseinpour, M.; Abdoos, H.; Alamdari, S.; Menéndez, J. L., Flexible nanocomposite scintillator detectors for medical applications: A review. *Sensors and Actuators A: Physical* **2024,** *378*, 115828.
12. Lowdon, M.; Martin, P. G.; Hubbard, M. W. J.; Taggart, M. P.; Connor, D. T.; Verbelen, Y.; Sellin, P. J.; Scott, T. B., Evaluation of Scintillator Detection Materials for Application within Airborne Environmental Radiation Monitoring. **2019,** *19* (18), 3828.
13. del Re, D., Timing performance of the CMS ECAL and prospects for the future. *J. Phys. Conf. Ser.* **2015,** *587* (1), 012003.
14. Fabjan, C. W.; Gianotti, F., Calorimetry for particle physics. *Rev. Mod. Phys.* **2003,** *75* (4), 1243-1286.
15. Knoll, G. F., *Radiation detection and measurement*. John Wiley & Sons: 2010.
16. Conti, M., Focus on time-of-flight PET: the benefits of improved time resolution. *Eur. J. Nucl. Med. Mol. Imaging.* **2011,** *38* (6), 1147-1157.
17. Jones, T. L.; Townsend, D. W., History and future technical innovation in positron emission tomography. *Journal of medical imaging (Bellingham, Wash.)* **2017,** *4* (1), 011013-011013.
18. Gundacker, S.; Martinez Turtos, R.; Kratochwil, N.; Pots, R. H.; Paganoni, M.; Lecoq, P.; Auffray, E., Experimental time resolution limits of modern SiPMs and TOF-PET detectors exploring different scintillators and Cherenkov emission. *Phys. Med. Biol.* **2020,** *65* (2), 025001.
19. Lecoq, P.; Morel, C.; Prior, J. O.; Visvikis, D.; Gundacker, S.; Auffray, E.; Križan, P.; Turtos, R. M.; Thers, D.; Charbon, E.; Varela, J.; de La Taille, C.; Rivetti, A.; Breton, D.; Pratte, J.-F.; Nuyts, J.; Surti, S.; Vandenberghe, S.; Marsden, P.; Parodi, K.; Benlloch, J. M.; Benoit, M., Roadmap toward the 10 ps time-of-flight PET challenge. *Phys. Med. Biol.* **2020,** *65* (21), 21RM01.
20. Auffray, E.; Buganov, O.; Fedorov, A.; Korjik, M.; Lecoq, P.; Tamulaitis, G.; Tikhomirov, S.; Vasil'ev, A., New detecting techniques for a future calorimetry. *J. Phys. Conf. Ser.* **2015,** *587* (1), 012056.
21. Duarte, J. P. B. S.; Filho, L. M. d. A.; Filho, E. F. d. S.; Farias, P. C. M. A.; de Seixas, J. M., Online energy reconstruction for calorimeters under high pile-up conditions using deconvolutional techniques. *J. Instrum.* **2019,** *14* (12), P12017.

22. Ilie, S.; Schönbacher, H.; Tavlet, M., Radiation-damage measurements on PVT-based plastic scintillators. *Nucl. Phys. B Proc. Suppl.* **1993,** *32,* 384-391.
23. Papageorgakis, C.; Al-Sheikhly, M.; Belloni, A.; Edberg, T. K.; Eno, S. C.; Feng, Y.; Jeng, G.-Y.; Kahn, A.; Lai, Y.; McDonnell, T.; Mohammed, A.; Palmer, C.; Perez-Gokhale, R.; Ricci-Tam, F.; Yang, Z.; Yao, Y., Dose rate effects in radiation-induced changes to phenyl-based polymeric scintillators. *Nucl. Instrum. Methods Phys. Res. A.* **2022,** *1042,* 167445.
24. Anand, A.; Zaffalon, M. L.; Erroi, A.; Cova, F.; Carulli, F.; Brovelli, S., Advances in Perovskite Nanocrystals and Nanocomposites for Scintillation Applications. *ACS Energy Lett.* **2024,** *9* (3), 1261-1287.
25. Koshimizu, M., Composite scintillators based on polymers and inorganic nanoparticles. *Functional Materials Letters* **2020,** *13* (06), 2030003.
26. Shevelev, V. S.; Ishchenko, A. V.; Vanetsev, A. S.; Nagirnyi, V.; Omelkov, S. I., Ultrafast hybrid nanocomposite scintillators: A review. *J. Lumin.* **2022,** *242,* 118534.
27. Singh, P.; Dosovitskiy, G.; Bekenstein, Y., Bright Innovations: Review of Next-Generation Advances in Scintillator Engineering. *ACS Nano* **2024,** *18* (22), 14029-14049.
28. Antonelli, A.; Auffray, E.; Brovelli, S.; Bruni, F.; Campajola, M.; Carsi, S.; Carulli, F.; De Nardo, G.; Di Meco, E.; Diociaiuti, E. J. a. p. a., Development of nanocomposite scintillators for use in high-energy physics. *Nucl. Instrum. Methods Phys. Res. A.* **2024,** *1069* (169877).
29. Gandini, M.; Villa, I.; Beretta, M.; Gotti, C.; Imran, M.; Carulli, F.; Fantuzzi, E.; Sassi, M.; Zaffalon, M.; Brofferio, C.; Manna, L.; Beverina, L.; Vedda, A.; Fasoli, M.; Gironi, L.; Brovelli, S., Efficient, fast and reabsorption-free perovskite nanocrystal-based sensitized plastic scintillators. *Nat. Nanotechnol.* **2020,** *15* (6), 462-468.
30. Koshimizu, M., Composite Scintillators. In *Plastic Scintillators: Chemistry and Applications*, Hamel, M., Ed. Springer International Publishing: Cham, 2021; pp 201-222.
31. Zaffalon, M. L.; Cova, F.; Liu, M.; Cemmi, A.; Di Sarcina, I.; Rossi, F.; Carulli, F.; Erroi, A.; Rodà, C.; Perego, J.; Comotti, A.; Fasoli, M.; Meinardi, F.; Li, L.; Vedda, A.; Brovelli, S., Extreme γ-ray radiation hardness and high scintillation yield in perovskite nanocrystals. *Nat. Photonics* **2022,** *16* (12), 860-868.
32. Erroi, A.; Mecca, S.; Zaffalon, M. L.; Frank, I.; Carulli, F.; Cemmi, A.; Di Sarcina, I.; Debellis, D.; Rossi, F.; Cova, F.; Pauwels, K.; Mauri, M.; Perego, J.; Pinchetti, V.; Comotti, A.; Meinardi, F.; Vedda, A.; Auffray, E.; Beverina, L.; Brovelli, S., Ultrafast and Radiation-Hard Lead Halide Perovskite Nanocomposite Scintillators. *ACS Energy Lett.* **2023,** *8* (9), 3883-3894.
33. Erroi, A.; Carulli, F.; Cova, F.; Frank, I.; Zaffalon, M. L.; Llusar, J.; Mecca, S.; Cemmi, A.; Di Sarcina, I.; Rossi, F.; Beverina, L.; Meinardi, F.; Infante, I.; Auffray, E.; Brovelli, S., Ultrafast Nanocomposite Scintillators Based on Cd-Enhanced $CsPbCl_3$ Nanocrystals in Polymer Matrix. *ACS Energy Lett.* **2024,** *9* (5), 2333-2342.
34. Anand, A.; Zaffalon, M. L.; Cova, F.; Pinchetti, V.; Khan, A. H.; Carulli, F.; Brescia, R.; Meinardi, F.; Moreels, I.; Brovelli, S., Optical and Scintillation Properties of Record-Efficiency CdTe Nanoplatelets toward Radiation Detection Applications. *Nano Lett.* **2022,** *22* (22), 8900-8907.
35. Carulli, F.; Cova, F.; Gironi, L.; Meinardi, F.; Vedda, A.; Brovelli, S., Stokes Shift Engineered Mn:CdZnS/ZnS Nanocrystals as Reabsorption-Free Nanoscintillators in High Loading Polymer Composites. *Adv. Opt. Mater.* **2022,** *10* (13), 2200419.
36. Yu, H.; Winardi, I.; Han, Z.; Prout, D.; Chatziioannou, A.; Pei, Q., Fast Spectroscopic Gamma Scintillation Using Hafnium Oxide Nanoparticles–Plastic Nanocomposites. *Chem. Mater.* **2024,** *36* (1), 533-540.
37. Winardi, I.; Han, Z.; Yu, H.; Surabhi, P.; Pei, Q., Nanocomposite Scintillators Loaded With Hafnium Oxide and Phosphorescent Host and Guest for Gamma Spectroscopy. *Chem. Mater.* **2024,** *36* (10), 5257-5263.
38. Kakavelakis, G.; Gedda, M.; Panagiotopoulos, A.; Kymakis, E.; Anthopoulos, T. D.; Petridis, K., Metal Halide Perovskites for High-Energy Radiation Detection. *Adv. Sci.* **2020,** *7* (22), 2002098.
39. Chen, Q.; Wu, J.; Ou, X.; Huang, B.; Almutlaq, J.; Zhumekenov, A. A.; Guan, X.; Han, S.; Liang, L.; Yi, Z.; Li, J.; Xie, X.; Wang, Y.; Li, Y.; Fan, D.; Teh, D. B. L.; All, A. H.; Mohammed, O. F.; Bakr, O. M.; Wu, T.; Bettinelli, M.; Yang, H.; Huang, W.; Liu, X., All-inorganic perovskite nanocrystal scintillators. *Nature* **2018,** *561* (7721), 88-93.
40. Tong, J.; Wu, J.; Shen, W.; Zhang, Y.; Liu, Y.; Zhang, T.; Nie, S.; Deng, Z., Direct Hot-Injection Synthesis of Lead Halide Perovskite Nanocubes in Acrylic Monomers for Ultrastable and Bright Nanocrystal–Polymer Composite Films. *ACS Appl. Mater. Interfaces* **2019,** *11* (9), 9317-9325.

41. Chen, W.; Zhou, M.; Liu, Y.; Yu, X.; Pi, C.; Yang, Z.; Zhang, H.; Liu, Z.; Wang, T.; Qiu, J.; Yu, S. F.; Yang, Y.; Xu, X., All-Inorganic Perovskite Polymer–Ceramics for Flexible and Refreshable X-Ray Imaging. *Adv. Funct. Mater.* **2022,** *32* (2), 2107424.
42. Mecca, S.; Pallini, F.; Pinchetti, V.; Erroi, A.; Fappani, A.; Rossi, F.; Mattiello, S.; Vanacore, G. M.; Brovelli, S.; Beverina, L., Multigram-Scale Synthesis of Luminescent Cesium Lead Halide Perovskite Nanobricks for Plastic Scintillators. *ACS Appl. Nano Mater.* **2023,** *6* (11), 9436-9443.
43. Wei, S.; Yang, Y.; Kang, X.; Wang, L.; Huang, L.; Pan, D., Room-temperature and gram-scale synthesis of $CsPbX_3$ (X = Cl, Br, I) perovskite nanocrystals with 50–85% photoluminescence quantum yields. *Chem. Commun.* **2016,** *52* (45), 7265-7268.
44. Huang, H.; Li, Y.; Tong, Y.; Yao, E.-P.; Feil, M. W.; Richter, A. F.; Döblinger, M.; Rogach, A. L.; Feldmann, J.; Polavarapu, L., Spontaneous Crystallization of Perovskite Nanocrystals in Nonpolar Organic Solvents: A Versatile Approach for their Shape-Controlled Synthesis. *Angew. Chem. Int. Ed.* **2019,** *58* (46), 16558-16562.
45. Shulenberger, K. E.; Ashner, M. N.; Ha, S. K.; Krieg, F.; Kovalenko, M. V.; Tisdale, W. A.; Bawendi, M. G., Setting an Upper Bound to the Biexciton Binding Energy in $CsPbBr_3$ Perovskite Nanocrystals. *J. Phys. Chem. Lett.* **2019,** *10* (18), 5680-5686.
46. Ashner, M. N.; Shulenberger, K. E.; Krieg, F.; Powers, E. R.; Kovalenko, M. V.; Bawendi, M. G.; Tisdale, W. A., Size-Dependent Biexciton Spectrum in $CsPbBr_3$ Perovskite Nanocrystals. *ACS Energy Lett.* **2019,** *4* (11), 2639-2645.
47. Děcká, K.; Pagano, F.; Frank, I.; Kratochwil, N.; Mihóková, E.; Auffray, E.; Čuba, V., Timing performance of lead halide perovskite nanoscintillators embedded in a polystyrene matrix. *J. Mater. Chem. C* **2022,** *10* (35), 12836-12843.
48. Huang, H.; Bodnarchuk, M. I.; Kershaw, S. V.; Kovalenko, M. V.; Rogach, A. L., Lead Halide Perovskite Nanocrystals in the Research Spotlight: Stability and Defect Tolerance. *ACS Energy Lett.* **2017,** *2* (9), 2071-2083.
49. du Fossé, I.; Mulder, J. T.; Almeida, G.; Spruit, A. G. M.; Infante, I.; Grozema, F. C.; Houtepen, A. J., Limits of Defect Tolerance in Perovskite Nanocrystals: Effect of Local Electrostatic Potential on Trap States. *J. Am. Chem. Soc.* **2022,** *144* (25), 11059-11063.
50. Gao, L.; Li, Q.; Sun, J.-L.; Yan, Q., Gamma-Ray Irradiation Stability of Zero-Dimensional $Cs_3Cu_2I_5$ Metal Halide Scintillator Single Crystals. *J. Phys. Chem. Lett.* **2023,** *14* (5), 1165-1173.
51. Forde, A.; Kilin, D., Defect Tolerance Mechanism Revealed! Influence of Polaron Occupied Surface Trap States on $CsPbBr_3$ Nanocrystal Photoluminescence: Ab Initio Excited-State Dynamics. *J. Chem. Theory Comput.* **2021,** *17* (11), 7224-7236.
52. Yuan, X.; Hou, X.; Li, J.; Qu, C.; Zhang, W.; Zhao, J.; Li, H., Thermal degradation of luminescence in inorganic perovskite $CsPbBr_3$ nanocrystals. *Phys. Chem. Chem. Phys.* **2017,** *19* (13), 8934-8940.
53. Yang, D.; Li, X.; Zeng, H., Surface Chemistry of All Inorganic Halide Perovskite Nanocrystals: Passivation Mechanism and Stability. *Adv. Mater. Interfaces* **2018,** *5* (8), 1701662.
54. Diroll, B. T.; Nedelcu, G.; Kovalenko, M. V.; Schaller, R. D., High-Temperature Photoluminescence of $CsPbX_3$ (X = Cl, Br, I) Nanocrystals. *Adv. Funct. Mater.* **2017,** *27* (21), 1606750.
55. Hajagos, T. J.; Liu, C.; Cherepy, N. J.; Pei, Q., High-Z Sensitized Plastic Scintillators: A Review. *Adv. Mater.* **2018,** *30* (27), 1706956.
56. Brown, J. A.; Laplace, T. A.; Goldblum, B. L.; Manfredi, J. J.; Johnson, T. S.; Moretti, F.; Venkatraman, A., Absolute light yield of the EJ-204 plastic scintillator. *Nucl. Instrum. Methods Phys. Res. A.* **2023,** *1054*, 168397.
57. Meinardi, F.; Akkerman, Q. A.; Bruni, F.; Park, S.; Mauri, M.; Dang, Z.; Manna, L.; Brovelli, S., Doped Halide Perovskite Nanocrystals for Reabsorption-Free Luminescent Solar Concentrators. *ACS Energy Lett.* **2017,** *2* (10), 2368-2377.
58. Bellotti, V.; Carulli, F.; Mecca, S.; Zaffalon, M. L.; Erroi, A.; Catalano, F.; Boventi, M.; Infante, I.; Rossi, F.; Beverina, L.; Brovelli, S.; Simonutti, R., Perovskite Nanocrystals Initiate One-Step Oxygen Tolerant PET-RAFT Polymerization of Highly Loaded, Efficient Plastic Nanocomposites. *Adv. Funct. Mater.* **2024,** *n/a* (n/a), 2411319.
59. Wei, M.; de Arquer, F. P. G.; Walters, G.; Yang, Z.; Quan, L. N.; Kim, Y.; Sabatini, R.; Quintero-Bermudez, R.; Gao, L.; Fan, J. Z.; Fan, F.; Gold-Parker, A.; Toney, M. F.; Sargent, E. H., Ultrafast narrowband exciton routing within layered perovskite nanoplatelets enables low-loss luminescent solar concentrators. *Nat. Energy.* **2019,** *4* (3), 197-205.

60. Xin, Y.; Zhao, H.; Zhang, J., Highly Stable and Luminescent Perovskite–Polymer Composites from a Convenient and Universal Strategy. *ACS Appl. Mater. Interfaces* **2018,** *10* (5), 4971-4980.
61. Cova, F.; Erroi, A.; Zaffalon, M. L.; Cemmi, A.; Di Sarcina, I.; Perego, J.; Monguzzi, A.; Comotti, A.; Rossi, F.; Carulli, F.; Brovelli, S., Scintillation Properties of CsPbBr$_3$ Nanocrystals Prepared by Ligand-Assisted Reprecipitation and Dual Effect of Polyacrylate Encapsulation toward Scalable Ultrafast Radiation Detectors. *Nano Lett.* **2024,** *24* (3), 905-913.
62. Rainò, G.; Landuyt, A.; Krieg, F.; Bernasconi, C.; Ochsenbein, S. T.; Dirin, D. N.; Bodnarchuk, M. I.; Kovalenko, M. V., Underestimated Effect of a Polymer Matrix on the Light Emission of Single CsPbBr3 Nanocrystals. *Nano Lett.* **2019,** *19* (6), 3648-3653.
63. Pirman, T.; Ocepek, M.; Likozar, B., Radical Polymerization of Acrylates, Methacrylates, and Styrene: Biobased Approaches, Mechanism, Kinetics, Secondary Reactions, and Modeling. *Ind. Eng. Chem. Res.* **2021,** *60* (26), 9347-9367.
64. Pryor, W. A.; Coco, J. H., Computer Simulation of the Polymerization of Styrene. The Mechanism of Thermal Initiation and the Importance of Primary Radical Termination. *Macromolecules* **1970,** *3* (5), 500-508.
65. Dutta, A.; Behera, R. K.; Dutta, S. K.; Das Adhikari, S.; Pradhan, N., Annealing CsPbX$_3$ (X = Cl and Br) Perovskite Nanocrystals at High Reaction Temperatures: Phase Change and Its Prevention. *J. Phys. Chem. Lett.* **2018,** *9* (22), 6599-6604.
66. Liang, S.; Zhang, M.; He, S.; Tian, M.; Choi, W.; Lian, T.; Lin, Z., Metal halide perovskite nanorods with tailored dimensions, compositions and stabilities. *Nature Synthesis* **2023,** *2* (8), 719-728.
67. Liang, S.; He, S.; Zhang, M.; Yan, Y.; Jin, T.; Lian, T.; Lin, Z., Tailoring Charge Separation at Meticulously Engineered Conjugated Polymer/Perovskite Quantum Dot Interface for Photocatalyzing Atom Transfer Radical Polymerization. *J. Am. Chem. Soc.* **2022,** *144* (28), 12901-12914.
68. Yoon, Y. J.; Chang, Y.; Zhang, S.; Zhang, M.; Pan, S.; He, Y.; Lin, C. H.; Yu, S.; Chen, Y.; Wang, Z.; Ding, Y.; Jung, J.; Thadhani, N.; Tsukruk, V. V.; Kang, Z.; Lin, Z., Enabling Tailorable Optical Properties and Markedly Enhanced Stability of Perovskite Quantum Dots by Permanently Ligating with Polymer Hairs. *Adv. Mater.* **2019,** *31* (32), 1901602.
69. Hintermayr, V. A.; Lampe, C.; Löw, M.; Roemer, J.; Vanderlinden, W.; Gramlich, M.; Böhm, A. X.; Sattler, C.; Nickel, B.; Lohmüller, T.; Urban, A. S., Polymer Nanoreactors Shield Perovskite Nanocrystals from Degradation. *Nano Lett.* **2019,** *19* (8), 4928-4933.
70. Raja, S. N.; Bekenstein, Y.; Koc, M. A.; Fischer, S.; Zhang, D.; Lin, L.; Ritchie, R. O.; Yang, P.; Alivisatos, A. P., Encapsulation of Perovskite Nanocrystals into Macroscale Polymer Matrices: Enhanced Stability and Polarization. *ACS Appl. Mater. Interfaces* **2016,** *8* (51), 35523-35533.
71. Wang, B.; Li, P.; Zhou, Y.; Deng, Z.; Ouyang, X.; Xu, Q., Cs$_3$Cu$_2$I$_5$ Perovskite Nanoparticles in Polymer Matrix as Large-Area Scintillation Screen for High-Definition X-ray Imaging. *ACS Appl. Nano Mater.* **2022,** *5* (7), 9792-9798.
72. Liu, M.; Wan, Q.; Wang, H.; Carulli, F.; Sun, X.; Zheng, W.; Kong, L.; Zhang, Q.; Zhang, C.; Zhang, Q.; Brovelli, S.; Li, L., Suppression of temperature quenching in perovskite nanocrystals for efficient and thermally stable light-emitting diodes. *Nat. Photonics* **2021,** *15* (5), 379-385.
73. Seth, S.; Ahmed, T.; De, A.; Samanta, A., Tackling the Defects, Stability, and Photoluminescence of CsPbX$_3$ Perovskite Nanocrystals. *ACS Energy Lett.* **2019,** *4* (7), 1610-1618.
74. Bodnarchuk, M. I.; Boehme, S. C.; ten Brinck, S.; Bernasconi, C.; Shynkarenko, Y.; Krieg, F.; Widmer, R.; Aeschlimann, B.; Günther, D.; Kovalenko, M. V.; Infante, I., Rationalizing and Controlling the Surface Structure and Electronic Passivation of Cesium Lead Halide Nanocrystals. *ACS Energy Lett.* **2019,** *4* (1), 63-74.
75. ten Brinck, S.; Zaccaria, F.; Infante, I., Defects in Lead Halide Perovskite Nanocrystals: Analogies and (Many) Differences with the Bulk. *ACS Energy Lett.* **2019,** *4* (11), 2739-2747.
76. Rodà, C.; Fasoli, M.; Zaffalon, M. L.; Cova, F.; Pinchetti, V.; Shamsi, J.; Abdelhady, A. L.; Imran, M.; Meinardi, F.; Manna, L.; Vedda, A.; Brovelli, S., Understanding Thermal and A-Thermal Trapping Processes in Lead Halide Perovskites Towards Effective Radiation Detection Schemes. *Adv. Funct. Mater.* **2021,** *31* (43), 2104879.
77. Kazes, M.; Udayabhaskararao, T.; Dey, S.; Oron, D., Effect of Surface Ligands in Perovskite Nanocrystals: Extending in and Reaching out. *Acc. Chem. Res.* **2021,** *54* (6), 1409-1418.
78. Smock, S. R.; Chen, Y.; Rossini, A. J.; Brutchey, R. L., The Surface Chemistry and Structure of Colloidal Lead Halide Perovskite Nanocrystals. *Acc. Chem. Res.* **2021,** *54* (3), 707-718.

79. Ijaz, P.; Imran, M.; Soares, M. M.; Tolentino, H. C. N.; Martín-García, B.; Giannini, C.; Moreels, I.; Manna, L.; Krahne, R., Composition-, Size-, and Surface Functionalization-Dependent Optical Properties of Lead Bromide Perovskite Nanocrystals. *J. Phys. Chem. Lett.* **2020,** *11* (6), 2079-2085.
80. Zito, J.; Infante, I., The Future of Ligand Engineering in Colloidal Semiconductor Nanocrystals. *Acc. Chem. Res.* **2021,** *54* (7), 1555-1564.
81. Bulin, A.-L.; Vasil'ev, A.; Belsky, A.; Amans, D.; Ledoux, G.; Dujardin, C., Modelling energy deposition in nanoscintillators to predict the efficiency of the X-ray-induced photodynamic effect. *Nanoscale* **2015,** *7* (13), 5744-5751.
82. Fratelli, A.; Zaffalon, M. L.; Mazzola, E.; Dirin, D.; Cherniukh, I.; Martínez, C. O.; Salomoni, M.; Carulli, F.; Meinardi, F.; Gironi, L.; Manna, L.; Kovalenko, M. V.; Brovelli, S., Size-dependent multiexciton dynamics governs scintillation from perovskite quantum dots. *arXiv, 25 Sep 2024, doi.org/10.48550/arXiv.2409.16994* (accessed 2024/11/19)
83. Baccaro, S.; Cemmi, A.; Di Sarcina, I.; Ferrara, G. J. F.; Safety, T. f. N.; Security Department Casaccia Research Centre, E., Gamma irradiation Calliope facility at ENEA-Casaccia Research Centre (Rome, Italy). **2019**, 49.
84. Yeon, Y.-H.; Shim, H.-E.; Park, J.-H.; Lee, N.-H.; Park, J.-Y.; Chae, M.-S.; Mun, J.-H.; Lee, J.-H.; Gwon, H.-J., Evaluation of Radiation Resistance of Polystyrene Using Molecular Dynamics Simulation. *Materials* **2022,** *15* (1), 346.

# Supporting Information

## Resurfaced CsPbBr$_3$ Nanocrystals Enable Free Radical Thermal Polymerization of Efficient Ultrafast Polyvinyl Styrene Nanocomposite Scintillators

Francesco Carulli*, Andrea Erroi, Francesco Bruni, Matteo L. Zaffalon, Mingming Liu, Roberta Pascazio, Abdessamad El Adel, Federico Catalano, Alessia Cemmi, Ilaria Di Sarcina, Francesca Rossi, Laura Lazzarini, Daniela E. Manno, Ivan Infante, Liang Li, and Sergio Brovelli*

*francesco.carulli@unimib.it
*sergio.brovelli@unimib.it

**Methods**

- *Chemicals:* Cesium carbonate ($Cs_2CO_3$, 99%), lead bromide ($PbBr_2$, 98%), sodium fluoride (NaF, 99%), didodecyl dimethylammonium bromide (DDABr, 98%), 1-octadecene (90%), oleylamine (98%), toluene (99%), ultrapure water (suitable for HPLC), methyl acetate (MeOAc, 98%), oleic acid (90%), vinyltoluene (VT, 96%), divinylbenzene (DVB, 80%) and lauryl peroxide (thermal polymer initiator, 98%) was purchased from Sigma-Aldrich. All chemical reagents were used directly and without further purification.

- *Preparation of Cs-oleate precursor*: To prepare the Cs-oleate precursor, 10 mmol of $Cs_2CO_3$ (3.258 g) was degassed in 40 mL of oleic acid and 1-octadecene mixture (1:1 by volume) at 120° C for 1 h under vacuum. The flask was then filled with argon and heated to 150° C until a clear mixture was obtained, subsequently cooled to room temperature for storage.

- *Synthesis of pristine $CsPbBr_3$ nanocrystals (NCs)*: To prepare pristine $CsPbBr_3$ NCs, a solution of 2 mmol $PbBr_2$ (0.734 g) in 1-octadecene (20 mL), oleic acid (5 mL) and oleylamine (5 mL) was degassed under vacuum at 120° C for 1 hour. The mixture was then heated to 180° C under an argon atmosphere until there was complete dissolution of the $PbBr_2$. 1 mL of the previously heated Cs-oleate solution was swiftly injected into the flask, allowed to react for 10 s and then cooled down in an ice bath. The product was recovered by centrifugation with MeOAc and redispersed in toluene, resulting in a stock solution of pristine $CsPbBr_3$ NCs.

- *Preparation of DDAF precursor*: 2.5 mmol DDABr (1.1566 g) and 2.5 mmol NaF (105 mg) were dissolved in toluene (25 ml) and ultrapure water (25 ml) respectively. To completely exchange the bromide anions for fluoride anions, these solutions were mixed under sonication for 30 minutes. The milky looking mixture was then centrifuged and the toluene solution containing the DDAF precursor (0.015 mmol ml$^{-1}$) was obtained.

- *DDAF-treatment on pristine $CsPbBr_3$ NCs*: A certain amount of DDAF precursor solution was added to the $CsPbBr_3$ stock solution (1 ml) with stirring for 30 min at room temperature. The mixture was centrifuged with MeOAc and the sediment redissolved in toluene. After a further centrifugation step, the supernatant was collected as the DDAF-treated $CsPbBr_3$ NCs solution.

- *Nanocomposites fabrication*: VT and DVB were purified from polymerization inhibitor by column extraction with basic alumina. A stock solution of VT/DVB (90/10% V/V) was prepared and stored in a freezer. Pristine and DDAF-treated $CsPbBr_3$ NCs solution was dried under nitrogen flow. After complete removal of the original solvent from the NCs, a variable amount of NCs powder (depending on the desired NCs concentration of the final nanocomposite) was added to the VT/DVB solution. The resulting monomer/NCs solution was first sonicated at room temperature to dissolve any possible NCs aggregation, followed by stirring for 30 min. Finally, 1000 ppm of thermal polymer initiator was added to the clear monomer/NCs solution while stirring, which was then transferred to a cylindrical glass mould, refilled with nitrogen by continuous flowing for 5 min to remove oxygen and then accurately sealed. To obtain uniform, non-scattering nanocomposite, polymerization was carried out by placing the mould in an oil bath at 65° C for 12 hours. After complete polymerization, the nanocomposite was allowed to cool to room temperature before extraction from the mould.

- *Transmission electron microscopy*: High Angle Annular Dark Field (HAADF) images in scanning TEM (STEM) mode and the high-resolution TEM images (HRTEM) of powdered NCs were acquired in a JEOL JEM 2200FS transmission electron microscope equipped with in-column Omega filter, operating at 200 kV and coupled with an EDX Oxford Xplore silicon drift detector, with 80 mm$^2$ effective area. Some selected samples have been studied with the same methods in a Holographic Jeol JEM-ARM200F NEOARM, double corrected for spherical aberration, equipped with a 4k Gatan CCD camera, at University of Salento. The TEM image on nanocomposite sample were collected in bright-field (BF) mode. The samples were sectioned with an ultramicrotome (Ultracut EM UC6, Leica) equipped with a diamond knife (Diatome) and sections of 70 nm in thickness were collected on a Cu TEM grid, 200 hexagonal mesh. The grid was coated with a thin layer of carbon to preserve the nanocomposite sections and avoid charging problems. BF-TEM imaging was carried out with a JEOL JEM 1011 transmission electron microscope operated at 100 kV.

- *Optical measurements*: All the spectroscopic characterization on NCs in toluene solution was performed using 1 mm optical quartz cuvettes. Absorption spectra and PL on nanocomposite were collecting after the samples were reduced to a thickness of 1 mm and then polished on both flat surfaces. Optical absorption measurements were performed with a Cary 50 UV–VIS spectrophotometer at normal incidence in dual beam mode with a spectral resolution of 0.5 nm. Steady-state PL measurements were performed exciting the samples with a 405 nm pulsed diode laser (Edinburgh Inst. EPL 405, 40 ps pulse width) and collecting emission light with a TM-C10083CA Hamamatsu Mini-Spectrometer. Time-resolved PL measurements were performed using a 3.06 eV (405 nm) picosecond pulsed laser diode (Picoquant LDH-P series, ~70 ps pulses) and collecting the emission with a phototube coupled to a Cornerstone 260 1/4 m visible–near-infrared monochromator (ORIEL) and a time-correlated single-photon counting unit (time resolution, ~400 ps). PL quantum yield measurements were performed on both solution and nanocomposites using an integrating sphere coupled to a spectrometer, a charge-coupled device and using a 2.62 eV (473 nm) continuous laser as excitation source. For this specific characterization, solution and nanocomposite were measured inside a sealed cylindrical glass vials (diameter 6 mm) using the exact same amount of material (total volume 0.3 mL). The equipment was previously calibrated using $10^{-5}$ M rhodamine 6G solution in ethanol as reference sample.

- *Radioluminescence measurements*: Steady-state RL measurements were performed by irradiating the samples at room temperature with a Philips 2274 X-ray tube with a tungsten target, equipped with a beryllium window and operated at 20 kV and 20 mA without filtering beam. At this voltage, X-rays are generated by the bremsstrahlung mechanism. Collection of RL spectra were recorded using a homemade equipment featuring a liquid nitrogen-cooled charge-coupled device (CCD, Jobin-Yvon Symphony II) coupled to a monochromator (Jobin-Yvon Triax 180) with 100 grooves/mm and 300 grooves/mm gratings as detection system. The spectra have been corrected for the setup optical response. The light yield was evaluated using a comparative method by measuring the RL signal of a commercial EJ276D plastic scintillator (with a LY of 8600 photons/MeV) with the same size and in the same experimental conditions of $CsPbBr_3$ nanocomposites.

- *Time-resolved scintillation experiments:* Samples were excited with a Hamamatsu XRT N5084 pulsed tungsten X-ray tube operating at 40 kV, where a PicoQuant PDL 800-B pulsed diode laser with 40 ps pulse width (full-width-at-half-maximum -FWHM) acts as excitation source of the X-ray tube. The energy spectrum of the produced X-rays ranges from 0 to 40 keV with a pronounced peak between 9 and 10 keV, due to tungsten L-characteristic X-ray, and mean energy of about 15 keV. The X-rays hit the sample after crossing a brass collimator. The scintillation light is collected in TCSPC by a Becker & Hickl HPM 100-07 hybrid photomultiplier tube (HPM). The signal of the HPM was processed by an ORTEC 9237 amplifier and timing discriminator and acted as stop signal at a Cronologic xTDC4 time-to-digital-converter (TDC). The start signal was given by the external trigger of the pulsed laser. The overall impulse response function (IRF) of the system was obtained as the analytical convolution between the measured IRF of the laser together with HPM and the IRF of the X-ray tube, resulting in around 160 ps FWHM. The RL was spectrally selected using an optical bandpass filter at 500 nm with 40 nm FWHM mounted to the HPM that removed parasitic contributions due to air excitation by X-rays.

- *γ-ray irradiation experiments*: Similar portions of the same nanocomposite sample were placed in individual polypropylene sealed vials whose gamma ray attenuation is negligible (Eppendorf Tubes). The vials were irradiated in a pool-type gamma irradiation chamber equipped with a $^{60}$Co (mean energy ~1.25 MeV) γ-source rods array, uniformly irradiating the composite at 3.05 kGy$_{AIR}$ h$^{-1}$ dose rate value. Nanocomposite samples were irradiated at different cumulated absorbed doses up to 1 MGy by varying the irradiation time. The irradiation has been carried out at the CALLIOPE Gamma Irradiation Facility at ENEA Casaccia Research Centre. Throughout the paper, the given dose is in air.

- *Correction of RL spectra for reabsorption losses:* The correction of the RL intensity for self-absorption was performed by comparing the spectrally integrated intensity of the tail-normalized RL spectra at the different [NC] values with the same extracted from the RL profile of highly diluted samples that are not subject to reabsorption. We note that this approach slightly overestimates the reabsorption losses, because for highly luminescent samples such as the NCs reported here, the high probability of re-emission of reabsorbed photons produces an "artificially" stronger red tail that the simple correction does not account for.

- *Mass attenuation coefficient calculation:* The mass attenuation coefficients ($\mu/\rho$) of the nanocomposites of the set under investigation were calculated from the NIST database (https://physics.nist.gov/PhysRefData/Xcom/html/xcom1.html) considering the chemical composition and the density of each component of the scintillator as $\mu/\rho_{NANOC} = \sum_i (\mu/\rho)_i \omega_i$ where $\omega_i$ is the fraction by weight of the component *i*. $\mu/\rho$ was evaluated at the energy corresponding to the maximum emission of the continuous X ray spectrum produced by a Bremsstrahlung mechanism, corresponding to approximately 7 KeV. The nanocomposite with the lowest $\mu/\rho$ (pure PVT) could attenuate above 93% of incident X ray radiation.

- *Molecular Dynamics (MD) Simulation:* To construct the NC model, we began by preparing a CsPbBr$_3$ NC with an edge length of 5 nm, where the surface is terminated with the CsBr layer. Surface Cs cations were selectively removed from the inorganic core, resulting in a charge-balanced Cs$_{1030}$Pb$_{1000}$Br$_{3030}$ structure. Using the compound attachment tool CAT software[1], we capped the NC with oleate and oleylammonium (OLA/OLAM) ligands, achieving 45% surface coverage for oleate and 47.3% for oleylammonium. In the case of the DDAF-treated NC, 36% of surface bromine atoms were replaced with fluorine, and 36% of DDA molecules were attached using the CAT software, in accordance with experimental data reported in a previous study[2]. Each NC model was then placed in a 22 nm cubic simulation box using PACKMOL[3], with 45,332 toluene molecules added as the coarse-grained solvent for both systems. To perform the MD simulations, we first conducted 5 ns canonical (NVT) sampling to equilibrate all systems (OLA-OLAM capped, DDAF capped) at constant volumes and temperatures of 300 K and 338 K respectively, to mimic both the room temperature and polymerization condition. As previously mentioned, toluene was used as the solvent in all cases and it was introduced through a coarse-grained model[4]. The NVT equilibration, with a time step of 1 fs, was followed by an additional 2 ns equilibration in an isothermal-isobaric (NPT) ensemble at 1 atm pressure, maintaining temperatures of 300 K and 338 K. During both NVT and NPT equilibration stages, the positions of the core atoms of the NC were constrained to facilitate gradual system equilibration. After the completion of these equilibration phases, the system is deemed to be well-equilibrated at the target temperature and pressure conditions, allowing us to release the position restraints and proceed to the production phase. A 50 ns MD simulation was then performed for the production run. All simulations were carried out using GROMACS version 2023.1[5-12]. Force field (FF) parameters for the oleate/oleylammonium-capped $CsPbBr_3$ perovskite were developed using the auto-FOX package[13, 14], based on data from previous work[15] (reported in **Table S1**). To obtain optimized molecular dynamics (MD) parameters for the DDAF-capped model, a parameter optimization for a single DDAF ligand was performed using the Auto-FOX workflow, relying on the same reference parameters established in the prior study. For the MD simulations, we applied the smooth particle mesh Euler (SPME) method with beta-Euler splines[16] Temperature control at 300 K and 338 K was maintained using a velocity-rescaling thermostat[17] while pressure at 1 atm was regulated by the extended-ensemble Parrinello-Rahman barostat[18]. A 1 nm short-range cutoff was applied for both Lennard-Jones and Coulombic interactions. The force field parameters for the OLA and OLAM ligand tails were obtained using MATCH[19], a toolkit for automated assignment of CHARMM-based[20] atom types and force field parameters through comparison with a chemical fragment database.

|  | Charges (NC Core) | | σ (NC Core) | | | Charges (Ligand Anchors) | | | σ Charges (Ligand Anchors) | |
|---|---|---|---|---|---|---|---|---|---|---|
|  | *Pristine* | *DDAF-capped* |  | *Pristine* | *DDAF-capped* |  | *Pristine* | *DDAF-capped* |  | *Pristine* | *DDAF-capped* |
| **Cs** | 0.6316 | 0.5912 | **Cs Cs** | 0.543 | 0.515 | **O2D2** | -0.4553 | *** | **Cs C2O3** | 0.483 | *** |
| **Pb** | 1.2632 | 1.1824 | **Cs Pb** | 0.436 | 0.39 | **C2O3** | 0.4001 | *** | **Pb C2O3** | 0.25 | *** |
| **Br** | -0.6316 | -0.5912 | **Cs Br** | 0.393 | 0.395 | **N3P3** | -0.1642 | *** | **Br C2O3** | 0.232 | *** |
| **F** | *** | -0.7027 | **Pb Br** | 0.33 | 0.33 | **HGP2** | 0.2167 | *** | **Cs O2D2** | 0.182 | *** |
|  |  |  | **Pb Pb** | 0.603 | 0.484 | **N3P0** | *** | -0.42197 | **Pb O2D2** | 0.288 | *** |
|  |  |  | **Br Br** | 0.378 | 0.426 | **C324** | *** | -0.07035 | **Br O2D2** | 0.36 | *** |
|  |  |  | **Cs F** | *** | 0.342 | **C334** | *** | -0.24612 | **Cs N3P3** | 0.441 | *** |
|  |  |  | **Pb F** | *** | 0.273 | **HGP5** | *** | 0.1758 | **Pb N3P3** | 0.383 | *** |
|  |  |  | **Br F** | *** | 0.313 |  |  |  | **Br N3P3** | 0.35 | *** |
|  |  |  | **F F** | *** | 0.315 |  |  |  | **Cs HGP2** | 0.419 | *** |
|  |  |  |  |  |  |  |  |  | **Pb HGP2** | 0.287 | *** |
|  |  |  |  |  |  |  |  |  | **Br HGP2** | 0.258 | *** |
|  |  |  |  |  |  |  |  |  | **Br N3P0** | *** | 0.351 |
|  |  |  |  |  |  |  |  |  | **Cs N3P0** | *** | 0.366 |
|  |  |  |  |  |  |  |  |  | **N3P0 Pb** | *** | 0.356 |
|  |  |  |  |  |  |  |  |  | **F N3P0** | *** | 0.315 |
|  |  |  |  |  |  |  |  |  | **Br HGP5** | *** | 0.269 |
|  |  |  |  |  |  |  |  |  | **HGP5 Pb** | *** | 0.326 |
|  |  |  |  |  |  |  |  |  | **F HGP5** | *** | 0.185 |
|  |  |  |  |  |  |  |  |  | **Cs HGP5** | *** | 0.264 |
|  |  |  |  |  |  |  |  |  | **C334 Pb** | *** | 0.375 |
|  |  |  |  |  |  |  |  |  | **Br C334** | *** | 0.336 |
|  |  |  |  |  |  |  |  |  | **C334 F** | *** | 0.182 |
|  |  |  |  |  |  |  |  |  | **Cs C334** | *** | 0.398 |

**Table S1:** Calculated Force-Field Parameters of the pristine and DDAF-treated CsPbBr$_3$ NCs (core and ligands) for MD simulation. Charges are provided in elementary charge units and distances (σ) are reported in nm.

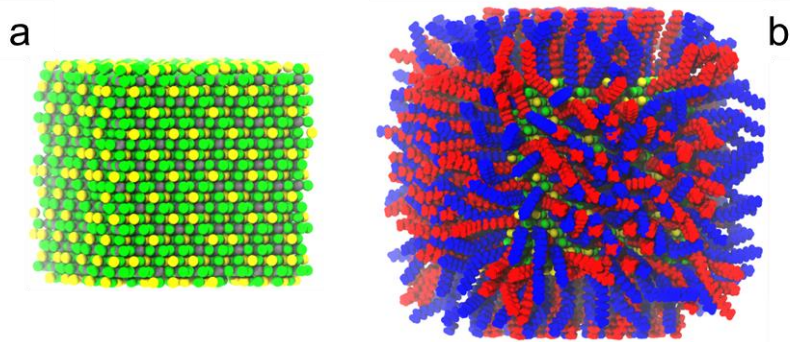

**Figure**: Sketches of NCs model used for the MD simulation: a) Structure of a core CsPbBr$_3$ NC (Cs: Yellow, Br: Green, Pb: Grey). b) Structure of a CsPbBr$_3$ NC capped with OLA (red ligands) and OLAM (blue ligands).

**Supporting Figures**

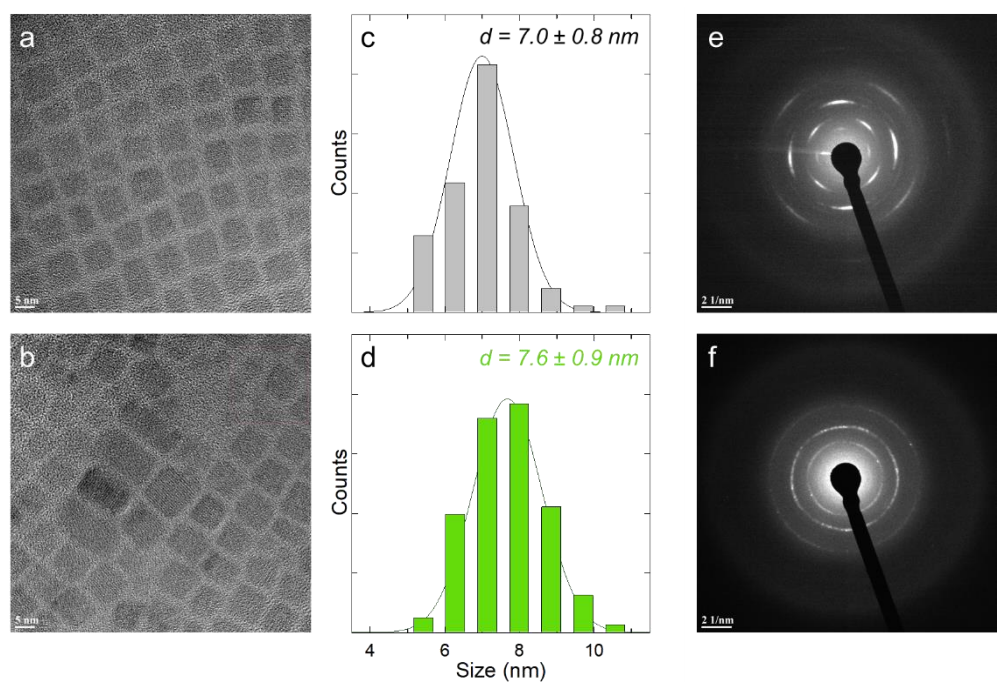

**Supporting Figure S1:** Representative HRTEM images of pristine (a) and DDAF-treated (b) $CsPbBr_3$ NCs. Size distribution of pristine (c, size=7.0±0.8 nm) and DDAF-treated (d, size=7.6±0.9 nm) $CsPbBr_3$ NCs calculated over 100 NCs. e, f) Diffraction patterns of pristine (e) and DDAF-treated (f) $CsPbBr_3$ NCs. Upon treatment there are no structural changes as evidenced by the exact same rings in the patterns. The difference in intensity along the rings derives from the different geometric arrangement of the NCs.

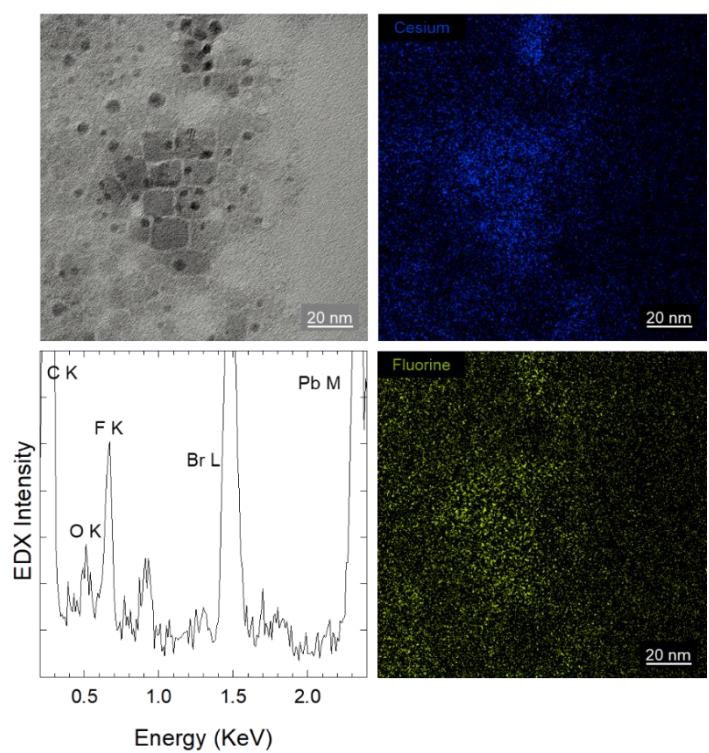

**Supporting Figure S2:** EDX maps of a DDAF-treated NCs samples, highlighting a higher fluorine signal in correspondence of NCs.

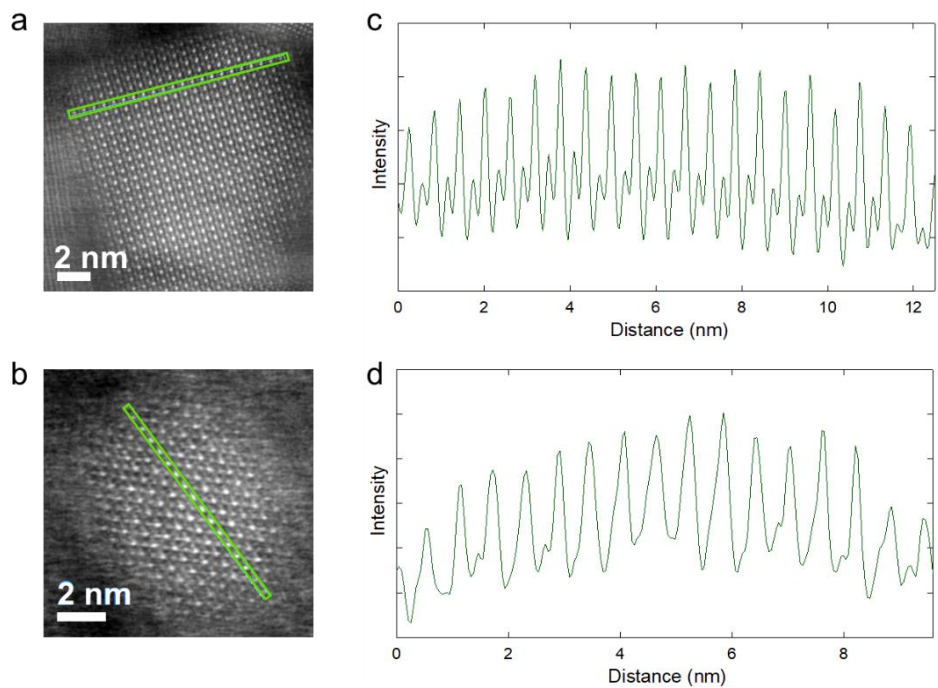

**Supporting Figure S3:** a, b) STEM-HAADF image of two representative DDAF-treated NCs. c, d) Line profile intensity along the small green rectangles highlighted in the a and b images, respectively. No significant variation in lattice parameter was found between the peripheral and central zones of the particles.

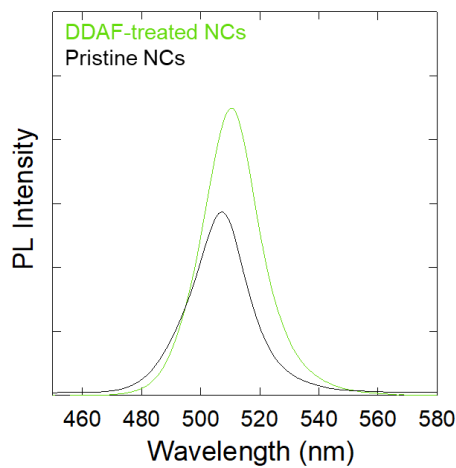

**Supporting Figure S4:** Comparative PL spectra of pristine (black curve, $\Phi_{PL}= 45\pm3\%$) and DDAF-treated (green curve, $\Phi_{PL}= 83\pm5\%$) $CsPbBr_3$ NCs in toluene solution.

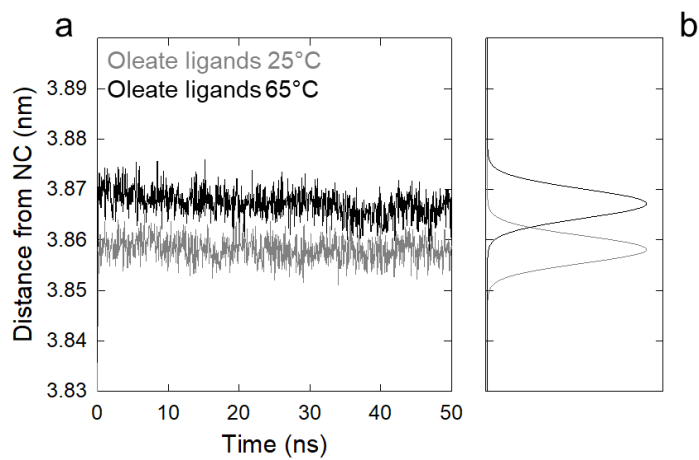

**Supporting Figure S5:** a) Time evolution of the average distances between the anchoring group of the oleate ligands and the center of mass a pristine $CsPbBr_3$ NC at 25°C (grey curve) and at 65°C (black curve). b) Time-mediated statistic of the data shown in plot a.

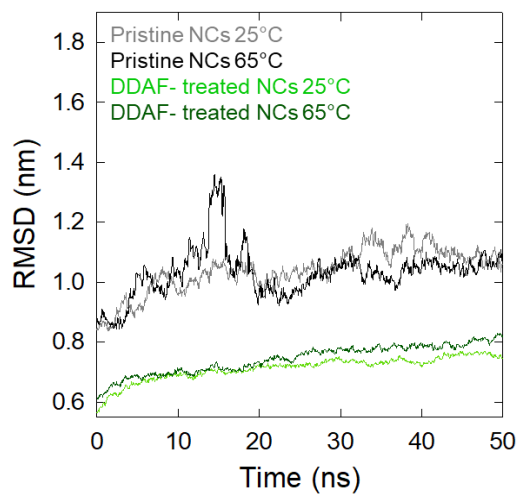

**Supporting Figure S6:** RMSD plot of the DDA or oleylamine ligand in DDAF-treated and pristine $CsPbBr_3$ NCs evaluated at 25°C (light green and grey curves, respectively) and at 65°C (dark green and black, respectively).

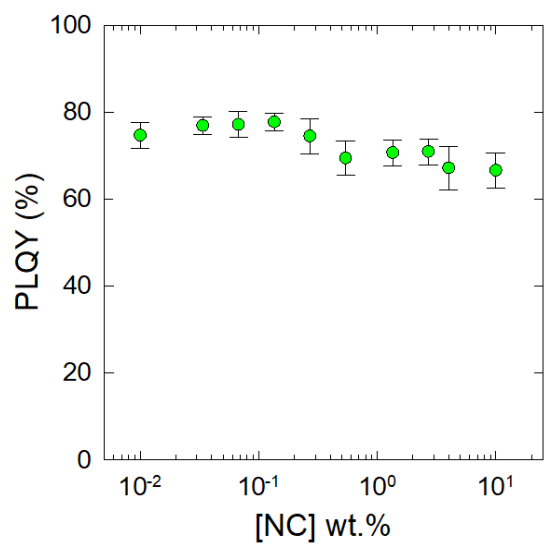

**Supporting Figure S7:** PLQY of PVT-based nanocomposites as a function of [NC] collected from slices of nanocomposites with decreasing thickness and excited with a 473 nm *cw* laser.

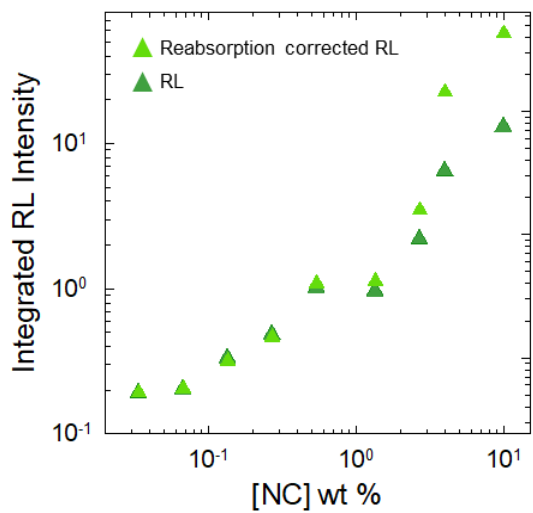

**Supporting Figure 8:** Spectrally integrated RL intensity as a function of [NC] before and after correcting the experimental data for the reabsorption

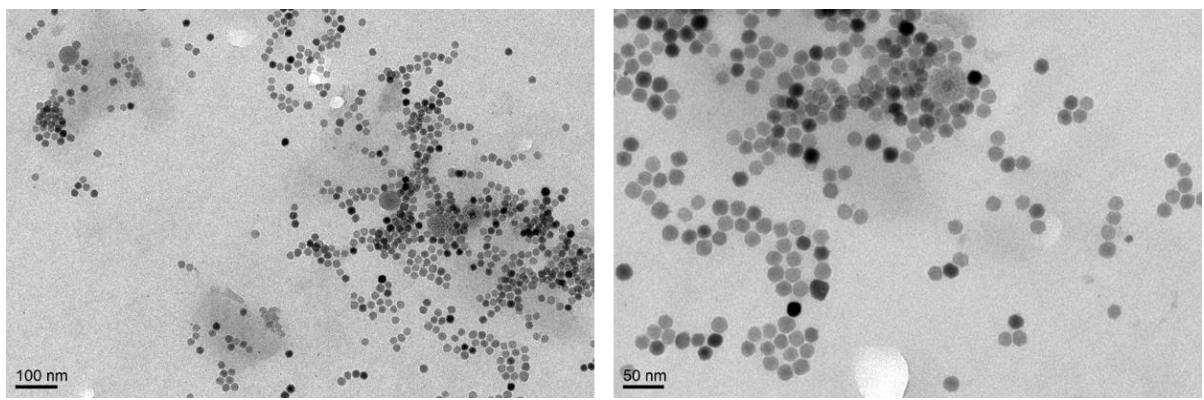

**Supplementary Figure S9:** TEM images of 70 nm thin slices of PVT composites containing [NC]=10 wt.%

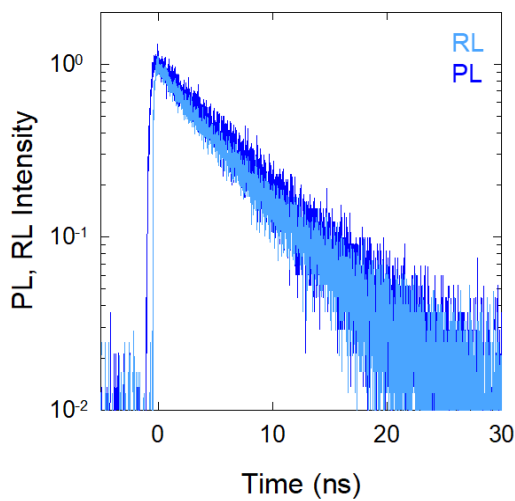

**Supporting Figure S10:** PL (excited with a 250nm pulsed laser) and RL dynamics of pure PVT sample.

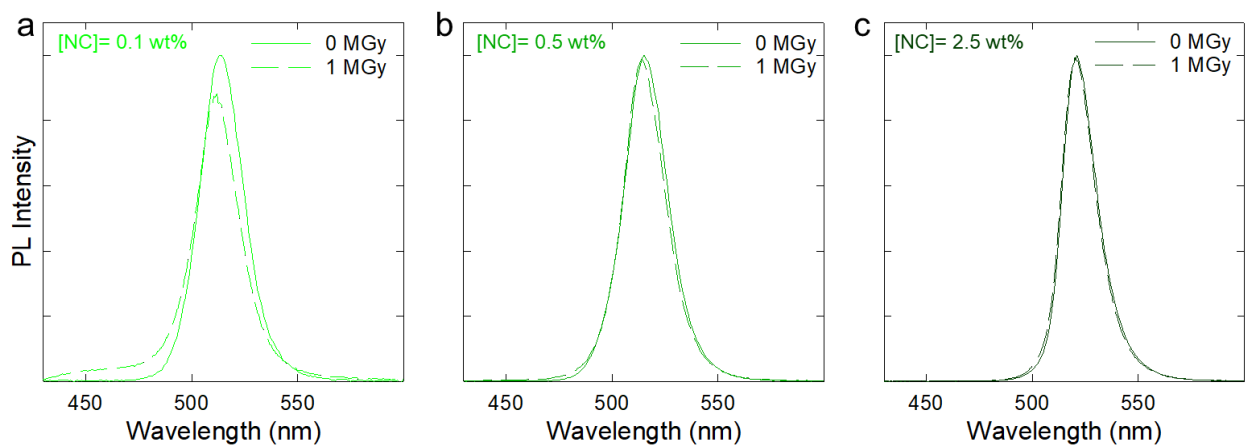

**Supporting Figure S11:** PL spectra of nanocomposite with [NC]= 0.1, 0.5 and 2.5 wt% before (continuous lines) and after (dashed lines) 1 MGy exposure. Samples were excited with a lamp at 420 nm, where the contribution of the radio-induced defect to nanocomposite absorption is negligible.

**Supporting References**

(1) van Beek, B.; Zito, J.; Visscher, L.; Infante, I., CAT: A Compound Attachment Tool for the Construction of Composite Chemical Compounds. *Journal of Chemical Information and Modeling* **2022,** *62* (22), 5525-5535.
(2) Liu, M.; Wan, Q.; Wang, H.; Carulli, F.; Sun, X.; Zheng, W.; Kong, L.; Zhang, Q.; Zhang, C.; Zhang, Q.; Brovelli, S.; Li, L., Suppression of temperature quenching in perovskite nanocrystals for efficient and thermally stable light-emitting diodes. *Nat. Photonics* **2021,** *15* (5), 379-385.
(3) Martínez, L.; Andrade, R.; Birgin, E. G.; Martínez, J. M., PACKMOL: A package for building initial configurations for molecular dynamics simulations. *Journal of Computational Chemistry* **2009,** *30* (13), 2157-2164.
(4) Marrink, S. J.; Risselada, H. J.; Yefimov, S.; Tieleman, D. P.; de Vries, A. H., The MARTINI Force Field: Coarse Grained Model for Biomolecular Simulations. *J. Phys. Chem. B* **2007,** *111* (27), 7812-7824.
(5) Abraham, M.; Alekseenko, A.; Bergh, C.; Blau, C.; Briand, E.; Doijade, M.; Fleischmann, S.; Gapsys, V.; Garg, G.; Gorelov, S.; Gouaillardet, G.; Gray, A.; Irrgang, M. E.; Jalalypour, F.; Jordan, J.; Junghans, C.; Kanduri, P.; Keller, S.; Kutzner, C.; Lemkul, J. A.; Lundborg, M.; Merz, P.; Miletić, V.; Morozov, D.; Páll, S.; Schulz, R.; Shirts, M.; Shvetsov, A.; Soproni, B.; Spoel, D. v. d.; Turner, P.; Uphoff, C.; Villa, A.; Wingbermühle, S.; Zhmurov, A.; Bauer, P.; Hess, B.; Lindahl, E.; Zenodo., E. L. G. M., GROMACS 2023.1 Manual Zenodo: 2023; p 852.
(6) Abraham, M. J.; Murtola, T.; Schulz, R.; Páll, S.; Smith, J. C.; Hess, B.; Lindahl, E., GROMACS: High performance molecular simulations through multi-level parallelism from laptops to supercomputers. *SoftwareX* **2015,** *1-2*, 19-25.
(7) Van Der Spoel, D.; Lindahl, E.; Hess, B.; Groenhof, G.; Mark, A. E.; Berendsen, H. J. C., GROMACS: Fast, flexible, and free. *Journal of Computational Chemistry* **2005,** *26* (16), 1701-1718.
(8) Páll, S.; Abraham, M. J.; Kutzner, C.; Hess, B.; Lindahl, E. In *Tackling Exascale Software Challenges in Molecular Dynamics Simulations with GROMACS*, Solving Software Challenges for Exascale, Cham, 2015//; Markidis, S.; Laure, E., Eds. Springer International Publishing: Cham, 2015; pp 3-27.
(9) Pronk, S.; Páll, S.; Schulz, R.; Larsson, P.; Bjelkmar, P.; Apostolov, R.; Shirts, M. R.; Smith, J. C.; Kasson, P. M.; van der Spoel, D.; Hess, B.; Lindahl, E., GROMACS 4.5: a high-throughput and highly parallel open source molecular simulation toolkit. *Bioinformatics* **2013,** *29* (7), 845-854.
(10) Lindahl, E.; Hess, B.; van der Spoel, D., GROMACS 3.0: a package for molecular simulation and trajectory analysis. *Molecular modeling annual* **2001,** *7* (8), 306-317.
(11) Berendsen, H. J. C.; van der Spoel, D.; van Drunen, R., GROMACS: A message-passing parallel molecular dynamics implementation. *Computer Physics Communications* **1995,** *91* (1), 43-56.
(12) Hess, B.; Kutzner, C.; van der Spoel, D.; Lindahl, E., GROMACS 4: Algorithms for Highly Efficient, Load-Balanced, and Scalable Molecular Simulation. *J. Chem. Theory Comput.* **2008,** *4* (3), 435-447.
(13) van Beek, B., Auto-FOX. 2023-05-10 ed.; Zenodo: 2023.
(14) Cosseddu, S.; Infante, I., Force Field Parametrization of Colloidal CdSe Nanocrystals Using an Adaptive Rate Monte Carlo Optimization Algorithm. *J. Chem. Theory Comput.* **2017,** *13* (1), 297-308.
(15) Pascazio, R.; Zaccaria, F.; van Beek, B.; Infante, I., Classical Force-Field Parameters for CsPbBr3 Perovskite Nanocrystals. *J. Phys. Chem. C* **2022,** *126* (23), 9898-9908.
(16) Essmann, U.; Perera, L.; Berkowitz, M. L.; Darden, T.; Lee, H.; Pedersen, L. G., A smooth particle mesh Ewald method. *J. Chem. Phys.* **1995,** *103* (19), 8577-8593.
(17) Bussi, G.; Donadio, D.; Parrinello, M., Canonical sampling through velocity rescaling. *J. Chem. Phys.* **2007,** *126* (1).
(18) Parrinello, M.; Rahman, A., Polymorphic transitions in single crystals: A new molecular dynamics method. *J. Appl. Phys.* **1981,** *52* (12), 7182-7190.
(19) Yesselman, J. D.; Price, D. J.; Knight, J. L.; Brooks Iii, C. L., MATCH: An atom-typing toolset for molecular mechanics force fields. *Journal of Computational Chemistry* **2012,** *33* (2), 189-202.
(20) Vanommeslaeghe, K.; Hatcher, E.; Acharya, C.; Kundu, S.; Zhong, S.; Shim, J.; Darian, E.; Guvench, O.; Lopes, P.; Vorobyov, I.; Mackerell Jr, A. D., CHARMM general force field: A force field for drug-like molecules compatible with the CHARMM all-atom additive biological force fields. *Journal of Computational Chemistry* **2010,** *31* (4), 671-690.